\documentclass[twocolumn]{aastex7}
\usepackage{amsmath}
\usepackage{wasysym}
\usepackage{graphicx}
\usepackage{amssymb}
\usepackage{epstopdf}
\usepackage{mathrsfs}
\usepackage{anyfontsize}
\usepackage{natbib}
\usepackage{color}
\usepackage{lipsum}
\usepackage{diagbox}

\newcommand{\Msun}{M_{\odot}}
\newcommand{\etash}{\eta_{\rm sh}}

\shorttitle{Gotta Go Fast}
\shortauthors{Paradiso \& Coughlin}
\begin{document} 
\title{Gotta Go Fast: A Generalization of the Escape Speed to Fluid-dynamical Explosions and Implications for Astrophysical Transients}
\author[0009-0003-8285-0702]{Daniel A.~Paradiso}
\email[show]{dparadis@syr.edu}
\affiliation{Department of Physics, Syracuse University, Syracuse, NY 13210, USA}
\author[0000-0003-3765-6401]{Eric R.~Coughlin}
\email[show]{ecoughli@syr.edu}
\affiliation{Department of Physics, Syracuse University, Syracuse, NY 13210, USA}


\begin{abstract}
A star's ability to explode in a core-collapse supernova is correlated with its density profile, $\rho(r)$, such that compact stars with shallow density profiles preferentially ``fail'' and produce black holes. This correlation can be understood from a mass perspective, as shallower density profiles enclose $\sim 3M_{\odot}$ (i.e., the maximum neutron-star mass) at relatively small radii, but could also be due to the fact that a shockwave (driving the explosion) inevitably stalls if the density profile into which it propagates is shallower than $\rho(r) \propto r^{-2}$. Here we show that this condition -- the density profile being steeper than $\rho \propto r^{-2}$ -- is necessary, but not sufficient, for generating a strong explosion. In particular, we find solutions to the fluid equations that describe a shockwave propagating at a fixed fraction of the local freefall speed into a temporally evolving, infalling medium, the density profile of which scales as $\rho \propto r^{-n}$ at large radii. The speed of the shock diverges as $n\rightarrow 2$ and declines (eventually to below the Keplerian escape speed) as $n$ increases, while the total energy contained in the explosion approaches zero as the shock recedes to large distances. These solutions therefore represent fluid-dynamical analogs of marginally bound orbits, and yield the ``shock escape speed'' as a function of the density profile. We also suggest that stellar explodability is correlated with the power-law index of the density at $\sim10^9$ cm, where the neutrino diffusion time equals the local dynamical time for most massive stars, which agrees with supernova simulations.
\end{abstract}


\section{Introduction}
\label{sec:intro}
Core-collapse supernovae (CCSNe) signify the deaths of some massive stars, and although the mechanism responsible for driving these luminous explosions remains uncertain (e.g., \citealt{Bethe85, Janka96, Mosta14, sukhbold16, Muller17, Oconnor18, Summa18, Couch20, Burrows24}), the generation of a shockwave from the formation and ``bounce'' of a neutron star remains a fundamental and likely necessary ingredient \citep{Arnett66}. When the energy of the shockwave far exceeds the binding energy of the star, gravitational effects are insignificant and the temporal evolution of the shockwave and the post-shock fluid can be understood analytically with the energy-conserving, Sedov-Taylor (ST) self-similar solution \citep{Sedov456, Taylor50, Chevalier76, Matzner99}. However, observational and theoretical evidence suggests that many massive stars do not produce successful explosions and -- provided that the star is not rapidly rotating at the onset of core-collapse, which could give rise to an accretion-powered explosion \citep{Woosley93, MacFadyen99} -- the shockwave stalls prior to reaching the stellar surface, resulting in a failed supernova (FSN) (e.g., \citealt{horiuchi11, OConnor11, smartt15, Ertl16, sukhbold16, Burrows24b}). 

Even before the shock stalls in a FSN, gravitational effects significantly modify the propagation of the shockwave once its speed approaches the freefall value. If the shock starts out in the energy-conserving/ST limit and the density of the gas into which the shock is advancing, $\rho$, is a power-law in radius $r$ (i.e., $\rho \propto r^{-n}$ with $n$ being a positive number), then energy conservation necessitates that the shock speed, $V$, varies with radius as $V\propto r^{\left(n-3\right)/2}$. The velocity of an energy-conserving shock therefore declines more rapidly with radius than the freefall speed ($\propto r^{-1/2}$) when $n<2$, which suggests that, provided enough time, the shock will inevitably decelerate to speeds comparable to the escape speed and subsequently stall. \citet{Paradiso24} recently developed an analytical solution for the specific case of an initially strong shock propagating into an infalling medium with $\rho \propto r^{-3/2}$, showing that when the energy of the explosion is below a critical value, the shockwave does indeed stall within the progenitor not long after it decelerates to near the escape speed. 

These arguments suggest that strong shockwaves (i.e., those with velocities significantly in excess of the stellar escape speed) are most readily produced by progenitors with steep density profiles, and this behavior \emph{is} typically observed in studies of the structure of the region in the immediate vicinity of the iron core of core-collapse progenitors \citep{Fryer99, OConnor11, Ertl16}. More specifically, the rate at which the density declines in the inner-region of supergiant stars has been shown to be correlated with a progenitor's potential to produce a successful explosion, with steeper density gradients resulting in a higher likelihood of explosion. One might therefore conclude that massive-star envelopes with density profiles steeper than $\propto r^{-2}$ -- which are actualized by some hydrogen-stripped stars (e.g., \citealt{Fernandez18, Woosley19, Ertl20, Vartanyan21}) -- always lead to successful explosions. However, while a density profile steeper than $\rho \propto r^{-2}$ is a \emph{necessary} condition for the shock to approach the strong limit, it may not be sufficient. 

In support of this possibility, \cite{coughlin18} showed that a shock Mach number must exceed a critical value in a hydrostatic envelope (which has $\rho \propto r^{-n}$ with $n>2$) to transition to the strong limit, and the self-similar solutions with this Mach number are unstable to radial perturbations \citep{coughlin19}, i.e., perturbations to the shock position and velocity grow with time and the solutions asymptotically approach the ST regime. Their investigation was focused on the dynamics of weak shocks, which form in outer regions of some CCSN progenitors as a response to the gravitational energy radiated away by neutrinos during neutron star formation \citep{Nadezhin80, Lovegrove13, Piro13, Lovegrove17, Coughlin18mefsn,Fernandez18}. Furthermore, their solutions predict that the fluid in the region interior to the shockwave accretes onto the central gravitating object, and are therefore applicable to FSNe, i.e., scenarios where the neutron star bounce shock is unable to drive a successful explosion and a black hole is formed. An apparent follow-up question from the work of \cite{coughlin18} and \cite{Paradiso24} is therefore: what constraints can be placed on the propagation of a shockwave in a non-hydrostatic, infalling medium with a density power-law index $n>2$ at asymptotically large radii, and is there similarly a new family of self-similar solutions that describe this scenario? Specifically, the case analyzed by \cite{Paradiso24} where the density profile $\rho\propto r^{-3/2}$, is a time-steady solution when the ambient medium is in freefall. Finding solutions where $n\geq2$ and the ambient medium is non-hydrostatic therefore requires the incorporation of the time-dependent nature of the ambient fluid.

In this paper, we show that there are indeed such self-similar solutions. These new solutions predict that the shock speed scales as a fixed fraction of the escape speed and describe the propagation of the shockwave and account for the time-dependent dynamics of the ambient medium. Similar to the solutions found by \citet{coughlin18} and analogous to the Bondi problem, a defining characteristic of these solutions is the existence of a sonic point within the post-shock flow (such solutions are commonly referred to as Type II similarity solutions \citealt{Sedov59}). At small radii interior to this sonic point, the fluid is in freefall and accretes onto the central object, and therefore these solutions are most relevant to sub-energetic and FSN where continued accretion onto the neutron star leads to the formation of a black hole.

In Section \ref{sec: interp} we begin by providing a physical motivation for our new solutions. We then derive the solution which describes the pressureless freefall of the ambient medium and show that there exists a unique self-similar solution to the fluid equations that describes the propagation of a shockwave into aforementioned ambient medium in Section \ref{sec: sol}. In Section \ref{sec: results} we discuss the results of these solutions and provide specific examples. We then compare our analytical predictions to 1D numerical simulations performed with the hydrodynamics code \textsc{flash} \citep{fryxell00} in Section \ref{sec: simulations}. In Section \ref{sec: discussion} we compare this work with previous efforts, discuss the future extensions, limitations, and astrophysical implications of this work, before summarizing in Section \ref{sec: sum}.

\section{Motivation and Physical Setup}
\label{sec: interp}
Once a $\sim$ Chandrasekhar-mass iron core forms within a supergiant star, the innermost ($\sim few \times1000$~km) regions collapse within a fraction of a second. The region of the star exterior to this (the ``ambient medium''), however, is causally unaware of this catastrophe occurring in its interior, as the loss of pressure support is communicated to the overlying stellar envelope by a rarefaction wave (RW) that travels at the local sound speed. Consecutive shells of ambient fluid thus fall inward, becoming highly supersonic as they do so. If the density profile of the stellar envelope is approximately a power-law (see, e.g., the bottom panel of Figure 3 in \citealt{sukhbold16}), then the RW propagates according to
\begin{equation}
    \label{R rw}
    R_{\rm w}\left(t\right)=R_{\rm w, 0}\left(1+\frac{3}{2}\sqrt{\frac{\gamma}{n+1}}\frac{\sqrt{GM}}{{R_{\rm w, 0}}^{3/2}}t\right)^{2/3},
\end{equation}
where $R_{\rm w, 0}$ is the location of the rarefaction wave at $t=0$, $\gamma$ is the adiabatic index of the gas, and $n$ is the power-law index of the stellar envelope (cf.~Section $6$ of \citealt{coughlin19}). There are also self-similar solutions for the infalling material, and in the limit that $r/R_{\rm w}(t) \ll 1$ are 
\begin{equation}
    v = -\sqrt{\frac{2GM}{r}}, \,\,\, \rho \propto R_{\rm w}(t)^{-n}\left(\frac{r}{R_{\rm w}}\right)^{-3/2}, \,\,\, p/(\rho v^2) \simeq 0, \label{gRW}
\end{equation}
where $v$ and $p$ are the velocity and pressure, implying that the gas approaches $\sim$ pressureless freefall at radii much smaller than the RW. 

\begin{figure}
  \includegraphics[width=0.45\textwidth]{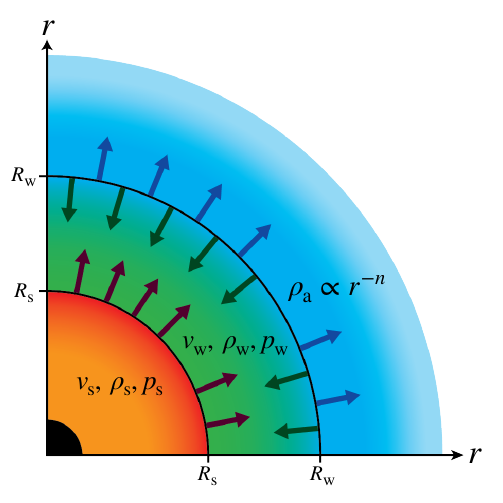}
  \caption{A depiction of RW ($R_{\rm w}$) and shock ($R_{\rm s}$) propagation in a CCSN, where the RW propagates into a $\rho \propto r^{-n}$, hydrostatic ambient medium (blue region) and causes shells of material to fall inward. At the same time, the shock is interior to the RW and advances into a now-infalling and time-dependent region (green) which, in the limit that $R_{\rm w} \gg R_{\rm s}$, is given by Equation \eqref{gRW}. }
  \label{fig:RW}
\end{figure}

While the RW propagates outward and causes successive shells of material to fall inward, the neutron star bounce shock is launched into the infalling gas described by Equation \eqref{gRW}. The shock tends to stall between $\sim 50-100$~km due to the dissociation of heavy nuclei and the ram pressure of the infalling envelope (e.g., \citealt{Bruenn85, Myra87, Baron90}), but can be subsequently ``revived'' by neutrino heating on timescales of $\sim 1$ s (the time taken for neutrinos to diffuse out of the neutron star; e.g., \citealt{burrows87}), leading to its continued and outward propagation (see diagram in Figure \ref{fig:RW}). By $\sim 1$ s, the RW has reached radii within the progenitor that have local dynamical times on the order of this timescale, being $few\times 10^{9}$ cm for typical supermassive stars (e.g., Figure 4 in \citealt{Fernandez18}), i.e., 2-3 orders of magnitude larger than the radius at which the shock initially stalls. For times $t \lesssim few$ seconds, the shock therefore sees (according to Equation \ref{gRW}) a static ambient medium that conforms to an $r^{-3/2}$ power-law scaling with density, meaning that it will decelerate in the gravitational field of the neutron star in accordance with the analytical solution of \cite{Paradiso24} and -- even if initially in the strong regime -- start to once again stall. However, after $\sim few$ seconds, the time dependence of the infalling gas is no longer ignorable, such that the density profile at a fixed radius $r$ declines with time as $\propto t^{1-2n/3}$, or $\propto R_{\rm w}^{-n}$ on surfaces that move at fixed fractions of $R_{\rm w}$. 

\begin{figure}
    \includegraphics[width=0.45\textwidth]{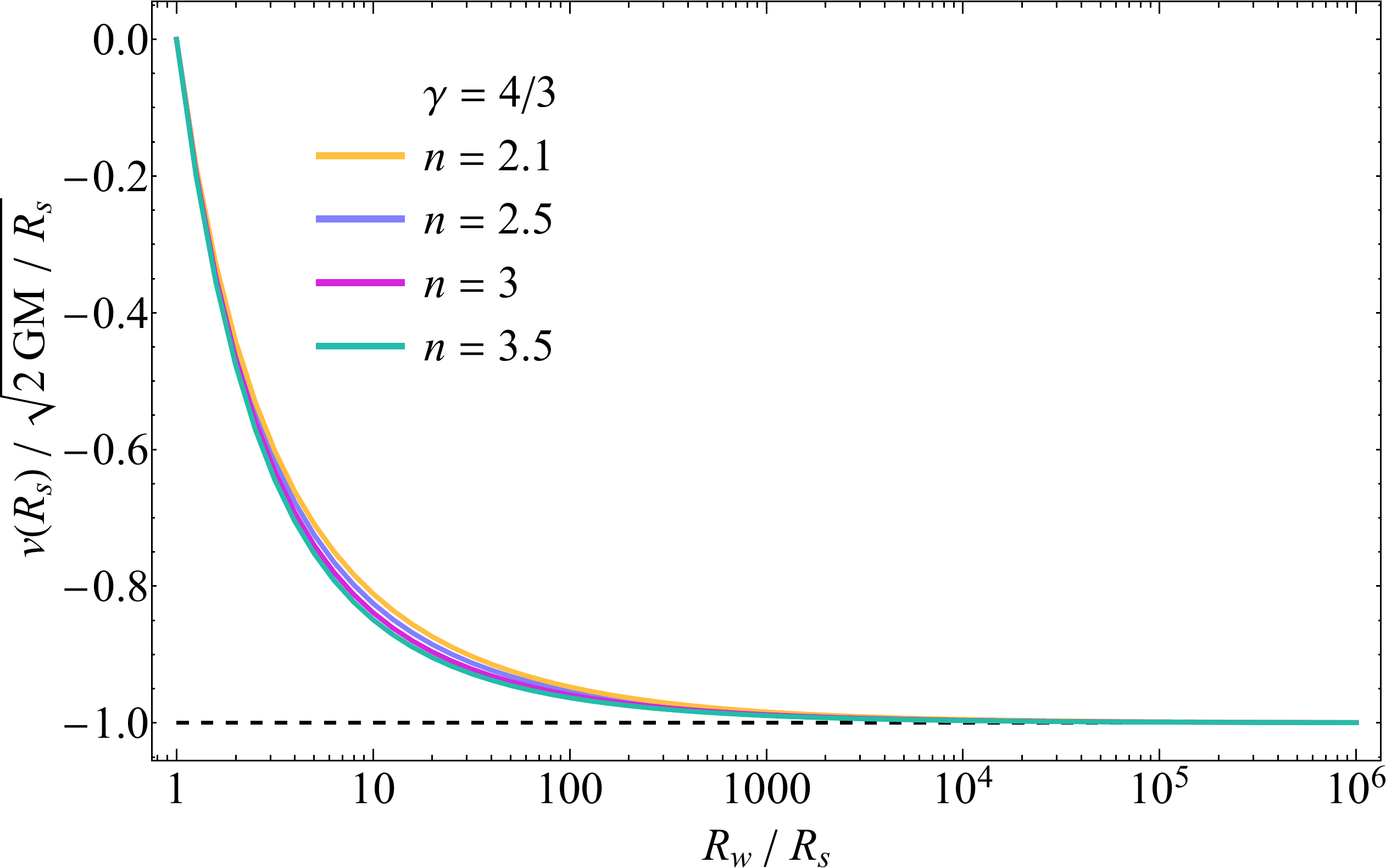}
    \caption{The ratio of the ambient velocity to the freefall speed at the location of the shockwave ($R_{\rm s}$) as a function of the ratio of the rarefaction wave radius ($R_{\rm w}$) to the the shockwave radius for different ambient density power-laws and adiabatic index $\gamma=4/3$. The black, dashed line shows when the absolute value of the velocity ratio is unity, and therefore when the ambient fluid is in freefall. For a given value of power-law index $n$, this is achieved when $R_{\rm w} \gg R_{\rm s}$. }
    \label{fig:v_RW}
\end{figure}

The preceding picture suggests that, during the early stages of CCSNe, the density of the medium into which the shock is advancing can be described by a simple power-law in radius with an overall time-dependent (and power-law) secular decline. However, Equation \eqref{gRW} is only accurate to leading order in $r/R_{\rm w}$, and the fluid impacting the shock will not be in complete freefall (see \citealt{Pejcha, Pochik24}, for example). Specifically, Figure \ref{fig:v_RW} shows the ratio of the ambient velocity to the freefall speed (from the \citealt{coughlin19} RW solution) as a function of the ratio of the RW radius and the shockwave radius (the latter denoted $R_{\rm s}$) for different envelope power-law indices and an adiabatic index of $\gamma=4/3$. It is apparent that differences arising from varying the power-law index are small, and that initial values of $R_{\rm w}/R_{\rm s}$ on the order of $10^{3}$ give infall speeds that are equal to the freefall value to within $\sim 1\%$. However, as the shock expands the fraction of freefall becomes significantly smaller than unity, which will have a non-negligible effect on the propagation of the shockwave. 

From the discussion in Section \ref{sec:intro}, we expect that stars with steep envelope density profiles (steeper than $\rho \propto r^{-2}$, specifically) could give rise to strong explosions, and the preceding discussion shows that a revived shock that moves outward at a fixed fraction of the freefall speed sees a temporal power-law decline in the ambient density. Owing to the lack of inherent spatial and temporal scales, one could conceivably find self-similar solutions in this regime that would represent fluid-dynamical analogs of marginally bound orbits. Simultaneously, however, the sub-freefall nature of the ambient gas (more realistically and as the shock moves to larger radii) could be important in modifying the propagation of the shock in ways that are not captured by using Equation \eqref{gRW} for the ambient fluid properties. 

To address both of these aspects of the problem -- the existence of self-similar solutions and the influence of the velocity of the ambient medium -- in the next section we first derive exact solutions to the fluid equations that describe the infall of gas onto an accreting object that are inherently time-dependent, but conform to power-laws in radius at asymptotically large distances; at these large distances the velocity of the gas is a fraction of freefall, where that fraction can be from zero (i.e., the gas is asymptotically at rest) to 1. We then show that there are self-similar solutions describing the propagation of a shockwave into these ambient profiles, such that the shock moves at a fixed fraction of the freefall speed. 

\section{Self-Similar Solutions}
\label{sec: sol}
\subsection{Ambient Solution}
\label{sec:ambient}
\begin{figure}
    \includegraphics[width=0.45\textwidth]{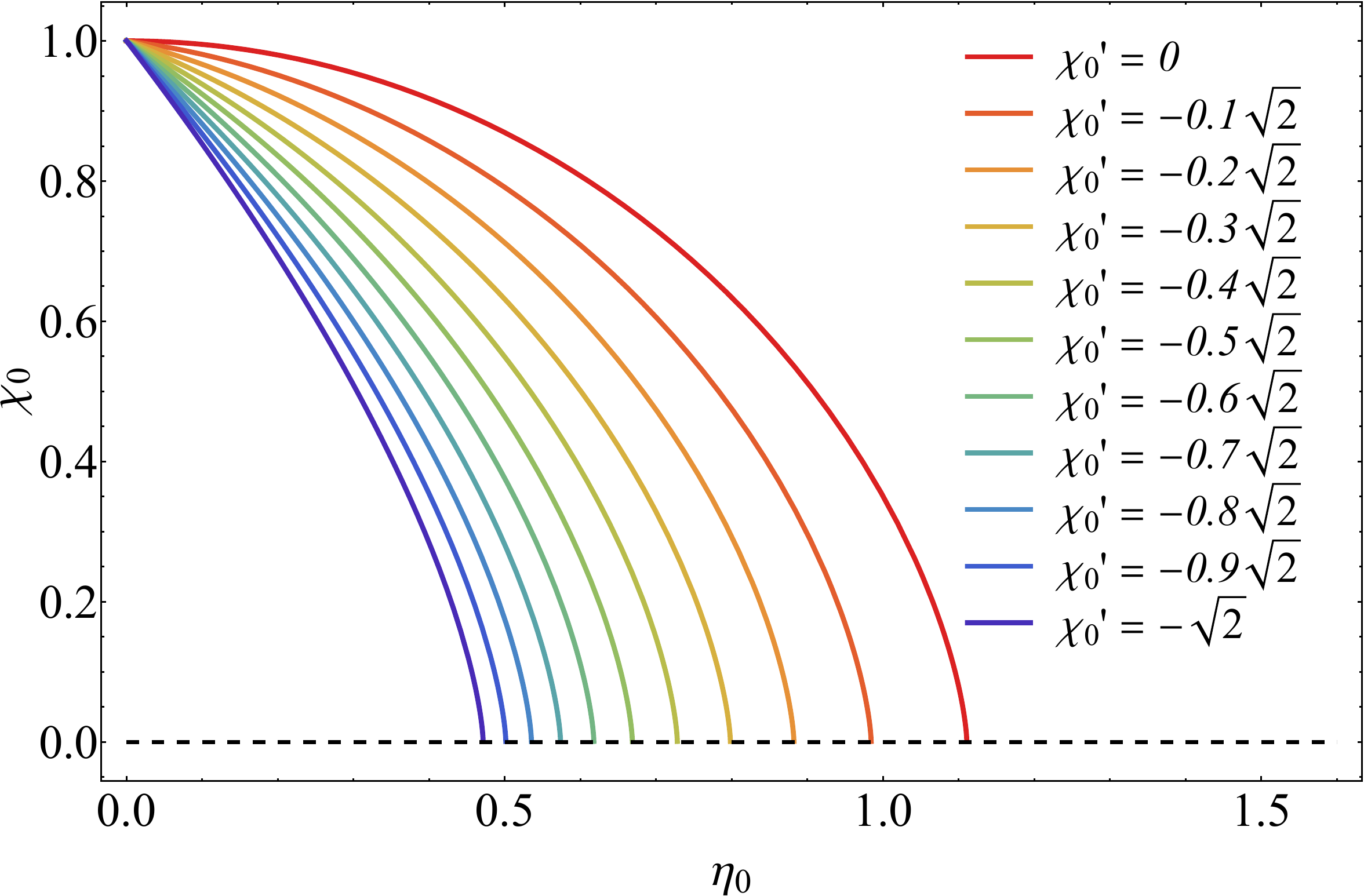}
    \caption{Solutions to Equation \eqref{chi_eta0_eom} with different initial velocity conditions ranging from an initially static ($\chi_0'=0$) fluid to one that is initially in freefall ($\chi_0'=-\sqrt{2}$). These solutions illustrate the characteristic timescale for a fluid element to fall back and accrete onto the central object.}
    \label{fig:chi_eta0}
\end{figure}

The Lagrangian equation of motion for a fluid element in pressureless freefall is given by
\begin{equation}
    \label{eom}
    \frac{\partial^2r}{\partial t^2}=-\frac{GM}{r^2},
\end{equation}
where $M$ is the mass of the central gravitating object, $r$ is the Lagrangian position of a fluid shell, and $\partial/\partial t$ is the Lagrangian time derivative. We parameterize the solution to this equation by defining
\begin{equation}
    \label{r}
    r=r_0 \chi_0\left(\eta_0\right),\quad \eta_0\equiv\frac{\sqrt{GM}t}{{r_0}^{3/2}},
\end{equation}
where $\eta_0$ is the dimensionless dynamical time in terms of the initial --- at $t=0$ --- Lagrangian position of the fluid element, $r_0$. Inserting Equation \eqref{r} into Equation \eqref{eom} then yields
\begin{equation}
    \label{chi_eta0_eom}
    \frac{\partial^2 \chi_0}{\partial{\eta_0}^2}=-\frac{1}{{\chi_0}^2}, 
\end{equation}
which, provided initial conditions, can be numerically integrated to solve for the functional form of $\chi_0\left(\eta_0\right)$. One such initial condition is $\chi_0(0) = 1$, while the assumption that $\chi_0$ depends only on $\eta_0$ restricts the second initial condition to $\chi_0'\left(0\right)=\chi_0'$, where $\chi'_0$ is a constant; $\chi'_0=0$ therefore describes an initially static fluid and $\chi_0'=-\sqrt{2}$ represents a fluid that is instantaneously in freefall. Since $\eta_0 \rightarrow 0$ in the limit that $r_0 \rightarrow \infty$, $\chi'_0$ also scales with the velocity of the fluid infinitely far from the central object. While in general Equation \eqref{chi_eta0_eom} must be numerically integrated to solve for $\chi_0\left(\eta_0\right)$, it is worth noting that an exact parametric solution to Equation \eqref{eom} exists when the fluid is initially at rest, and is given in, e.g., \citet{Spitzer78}. In terms of the variables $\chi_0$ and $\eta_0$, this solution is written as
\begin{equation}
    \label{exact param}
    \chi_0 = \frac{r}{r_0}=\cos^2\left(\beta\left(t\right)\right),
\end{equation}
where $\beta\left(t\right)$ satisfies
\begin{equation}
    \beta+\frac{1}{2}\sin\left(2\beta\right)=t\left(\frac{\sqrt{2GM}}{{r_0}^{3/2}}\right)=\sqrt{2}\eta_0.
\end{equation}
Differentiating Equation \eqref{exact param} therefore gives
\begin{equation}
    \chi'_0=-\sqrt{2}\tan\left(\beta\right) = 0
\end{equation}
since at $t=0$ ($\beta=0$) a given fluid element satisfies $r=r_0$. Figure \ref{fig:chi_eta0} shows a family of solutions to Equation \eqref{chi_eta0_eom} with initial conditions ranging from an initially static fluid to one that is initially in freefall. As expected, the timescale at which the fluid element reaches the central object is decreased by increasing its initial velocity. 

In the next section we will analyze the propagation of a shockwave into this infalling gas, and for this analysis it is useful to describe the ambient solutions in Eulerian coordinates. We therefore invert Equation \eqref{chi_eta0_eom} to write 
\begin{equation} 
\label{r0}
    r_0=r \chi\left(\eta\right), \quad \eta\equiv\frac{\sqrt{GM}t}{r^{3/2}}.
\end{equation}
Differentiating Equation \eqref{r0} with respect to time then shows
\begin{equation}
\label{amb v}
    v_{\rm a} = -\frac{\frac{d \chi}{d \eta}}{\chi-\frac{3}{2}\eta\frac{d \chi}{d \eta}}\sqrt{\frac{GM}{r}}\equiv - f_{\rm a}\left(\eta\right)\sqrt{\frac{GM}{r}},
\end{equation}
where $v_{\rm a}$ is the ambient velocity profile and in the last line we defined 
\begin{equation}
    f_{\rm a}\left(\eta\right) = \frac{\frac{d \chi}{d \eta}}{\chi-\frac{3}{2}\eta\frac{d \chi}{d \eta}}.
\end{equation}
Mass conservation gives 
\begin{equation}
\label{cont}
\rho\left(r,t\right)=\rho_0\left(r_0\right)\left(\frac{r_0}{r}\right)^{2}\left(\frac{\partial r_0}{\partial r}\right),
\end{equation}
where $\rho_0\left(r_0\right)$ is simultaneously the initial density profile and the density profile at large distances, which we write as a power-law: 
\begin{equation}
    \label{rho init}
    \rho_0=\rho_{\rm i} \left(\frac{r_0}{r_{\rm i}}\right)^{-n},
\end{equation}
with $\rho_{\rm i}$ being the density at the scale radius $r_{\rm i}$. Inserting Equation \eqref{r0} into Equation \eqref{cont} then gives
\begin{equation}
\label{amb dens}
\begin{split}
    \rho_{\rm a}\left(r,t\right)&=\rho_{\rm i}\left(\frac{r}{r_{\rm i}}\right)^{-n}{\chi}^{2-n}\left[\chi-\frac{3}{2}\eta \frac{d \chi}{d \eta}\right] \\     
    &\equiv \rho_{\rm i}\left(\frac{r}{r_{\rm i}}\right)^{-n}g_{\rm a}\left(n,\eta\right),
\end{split}
\end{equation}
where
\begin{equation}
    g_{\rm a}\left(n,\eta\right)={\chi}^{2-n}\left[\chi-\frac{3}{2}\eta \frac{d \chi}{d \eta}\right].
\end{equation}

\begin{figure*}
    \includegraphics[width=0.49\textwidth]{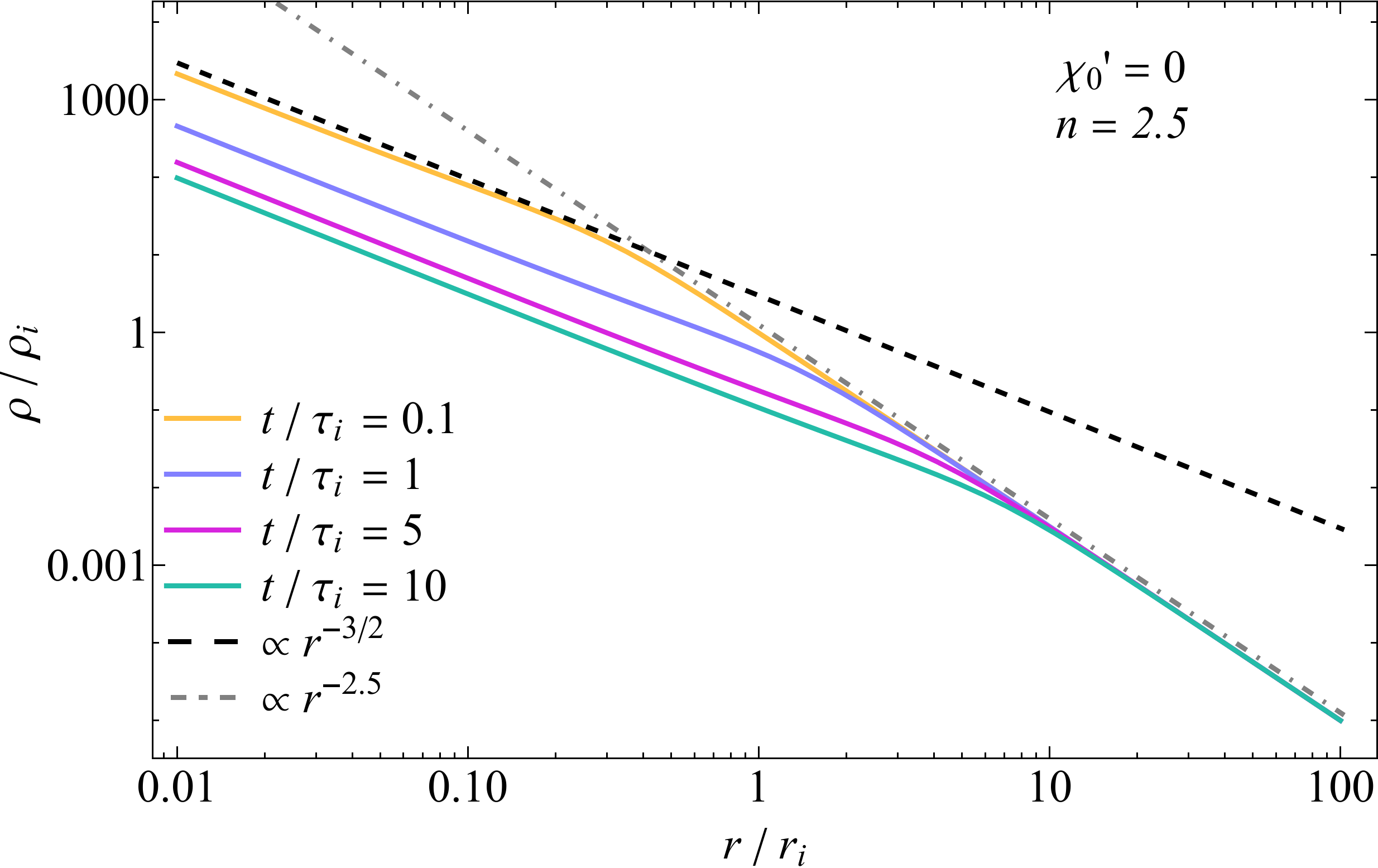}
    \includegraphics[width=0.49\textwidth]{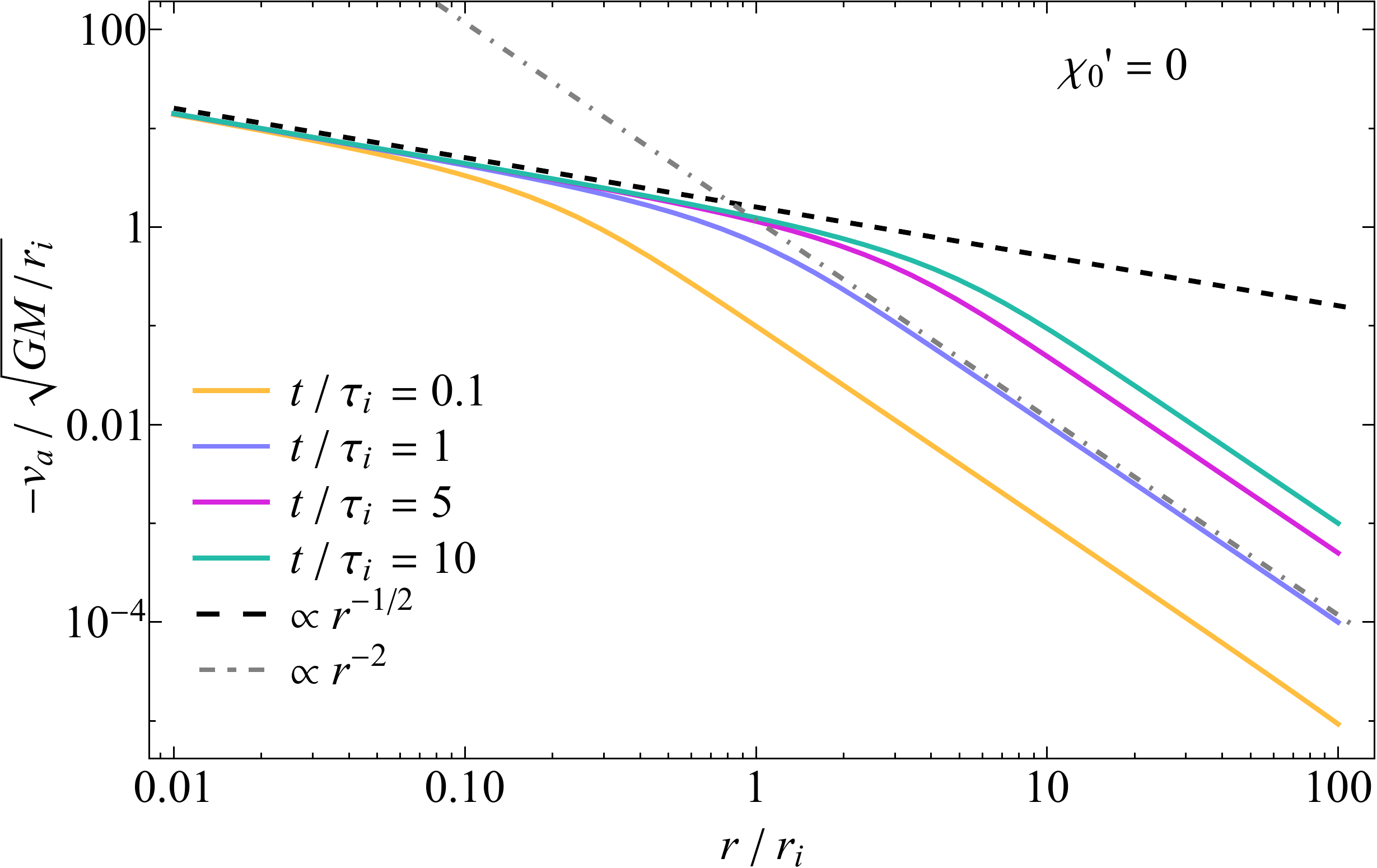}
    \caption{Left: ambient density profiles --- given by Equation \eqref{amb dens} --- as a function of radius for $\chi_0'=0$, $n=2.5$, and $\gamma = 4/3$ at the times in the legend. Here $\tau_{\rm i} = {r_{\rm i}}^{3/2}/\sqrt{GM}$ is the dynamical time at the scale radius $r_{\rm i}$. The black, dashed curve shows $\rho \propto r^{-2.5}$, to which each profile conforms at large radii, while the dot-dashed curve gives $\rho \propto r^{-3/2}$ and is the asmptotic scaling near the origin. Right: the (negative of the) ambient velocity normalized by $-\sqrt{GM/r_{\rm i}}$ as a function of radius for the times in the legend and $\chi_0'=0$ (note that the velocity is independent of the power-law index $n$). The black, dashed curve is $\propto r^{-1/2}$, showing that at small radii and late times (large $\eta$) the velocity approaches freefall, while at large radii and early times (small $\eta$) the velocity profiles are $\propto r^{-2}$ (gray, dot-dashed line).}
    \label{fig:amb_dens}
\end{figure*}
Equations \eqref{amb v} and \eqref{amb dens} are expressions that --- upon specifying values of $n$ and $\chi_0'$ --- provide exact, time-dependent solutions for the ambient medium. The left panel of Figure \ref{fig:amb_dens} shows the density for $n = 2.5$, and $\chi_0'=0$ normalized by $\rho_{\rm i}$ as a function of radius normalized by $r_{\rm i}$ at different times, given here in terms of the dynamical time at the initial scale radius, 
\begin{equation}
    \tau_{\rm i} = \frac{r_{\rm i}^{3/2}}{\sqrt{GM}}.
\end{equation} 
At small radii and late times --- i.e., for large values of $\eta$ --- the density profiles scale as the freefall scaling of $\rho \propto r^{-3/2}$, as shown by the black, dot-dashed curve. At large radii and early times (small $\eta$), however, the profiles are $\propto r^{-2.5}$, which is consistent with the above discussion, namely Equation \eqref{rho init}. In the right panel of Figure \ref{fig:amb_dens} we plot the ambient velocity normalized by $-\sqrt{GM/r_{\rm i}}$ as a function of radius normalized by $r_{\rm i}$ at the same times as the left panel for $\chi_0'=0$. The black, dashed curve shows the freefall scaling of $v \propto r^{-1/2}$, which these solutions approach at small radii and late times. Alternatively, at large radii and early times, the velocity profiles are $\propto r^{-2}$ --- as indicated by the gray, dot-dashed curve --- which is consistent with the the solution to Equation \eqref{eom} in the limit that $r$ has not changed substantially (i.e, $v \simeq -GMt/r^2$).

We note that Equation \eqref{chi_eta0_eom} can be integrated twice, and in most cases the result is not particularly enlightening as to the functional form of $\chi_0(\eta_0)$ and the corresponding spatial dependence and temporal evolution of the gas; an exception is when the gas is in instantaneous freefall ($\chi'_0 = -\sqrt{2}$), for which
\begin{equation}
    \chi_0 = \left(1-\frac{3}{\sqrt{2}}\eta_0\right)^{2/3} \quad \Rightarrow \quad v_{\rm a} = -\sqrt{\frac{2GM}{r}},
\end{equation}
i.e., the solution in Eulerian coordinates is time-steady in terms of the velocity. Although the density depends nontrivially on time for an arbitrary $r_0$, on any constant-$\eta$ surface the density appears to be a power-law in Eulerian radius. Therefore, self-similar solutions for which the shock moves out at a fixed fraction of the escape speed are, in this specific case, formally identical to those in which the ambient gas behaves according to Equation \eqref{gRW}, i.e., the small-$r$ limit of the RW solutions considered in \citet{coughlin19}. The instantaneous-freefall, ambient solution is therefore directly applicable and highly relevant to CCSNe in the limit that the RW is at a large radius relative to the shock, but the other solutions (i.e., those with $\chi'_0 \neq -\sqrt{2}$) measure the impact of the sub-freefall speed of the infalling gas. Moreover, the at-rest solution ($\chi'_0 = 0$) pertains to gas that is static at asymptotically large distances from the gravitating central object, and hence this case is qualitatively similar to CCSNe in that the RW is propagating into a hydrostatic medium. We therefore focus largely on these two limits --- $\chi'_0 = -\sqrt{2}$ and $\chi'_0 = 0$ --- in what follows.

\subsection{Self-similar Shock Solution}
\label{sec: shock}
The solutions for the ambient and infalling gas derived in the previous subsection are time-dependent, but we note that the time dependence is contained entirely in the single parameter $\eta$, being time relative to the local dynamical time. If a shock is now propagating into this medium, so that at any time the ambient profile holds only to radii outside of the shock radius $R_{\rm s}(t)$, then if the shock moves according to $R_{\rm s} \propto t^{2/3}$, the ambient density profile will be a power-law in time and the ambient velocity a fixed fraction of the local escape speed. This temporal scaling of the shock radius simultaneously implies that $\chi$ and $\chi'$ (or $\chi_0$ and $\chi_0'$ in terms of $\eta_0$) are constant at the location of the shock. We therefore seek solutions describing the propagation of a shockwave that satisfy 
\begin{equation}
    \label{RV shock}
    \frac{\sqrt{GM}t}{R_{\rm s}^{3/2}}=\eta_{\rm sh}, \quad \frac{dR_{\rm s}}{dt}\equiv V_{\rm s}=\frac{2}{3\eta_{\rm sh}}\sqrt{\frac{GM}{R_{\rm s}}}
\end{equation}
with $\eta_{\rm sh}$ constant, and write the velocity, density, and pressure as 
\begin{align}
    \label{param velo}
    v &= V_{\rm s}\left(t\right) f_{\rm s}\left(\xi\right), \\     
    \label{param dens}
    \rho &= \rho_{\rm i}{\left(\frac{R_{\rm s}\left(t\right)}{r_{\rm i}}\right)}^{-n}g_{\rm s}\left(\xi\right),  \\ 
    \label{param press}
    p &= \rho_{\rm i}{\left(\frac{R_{\rm s}\left(t\right)}{r_{\rm i}}\right)}^{-n}{V_{\rm s}\left(t\right)}^2h_{\rm s}\left(\xi\right),
\end{align}
where we have defined the dimensionless radial variable
\begin{equation}
    \xi=\frac{r}{R_{\rm s}\left(t\right)}.
\end{equation}

The spherically symmetric continuity, radial momentum, and entropy equations in Eulerian form are
\begin{align}
    &\frac{\partial{\rho}}{\partial{t}}+\frac{1}{r^2}\frac{\partial{}}{\partial{r}}\left[\rho r^2v\right]=0, \label{continuity}\\ 
    &\frac{\partial{v}}{\partial{t}}+v\frac{\partial{v}}{\partial{r}}+\frac{1}{\rho}\frac{\partial{p}}{\partial{r}}=-\frac{GM}{r^2}, \label{momentum}\\ 
    &\frac{\partial}{\partial{t}}\ln\left(\frac{p}{\rho^{\gamma}}\right)+v\frac{\partial}{\partial{r}}\ln\left(\frac{p}{\rho^{\gamma}}\right)=0. \label{energy eq}
\end{align}
Inserting Equations \eqref{param velo}-\eqref{param press} into Equations \eqref{continuity}-\eqref{energy eq} then yields the following, ordinary differential equations,
\begin{align}
    &-ng_{\rm s}-\xi \frac{d{g_{\rm s}}}{d{\xi}}+\frac{1}{{\xi}^2}\frac{d{}}{d{\xi}}\left[{\xi}^2f_{\rm s}g_{\rm s}\right]=0, \label{dim continuity}\\
    &-\frac{1}{2}f_{\rm s}+\left(f_{\rm s}-\xi\right)\frac{d{f_{\rm s}}}{d{\xi}}+\frac{1}{g_{\rm s}}\frac{d{h_{\rm s}}}{d{\xi}}=-\frac{9}{4}{\eta_{\rm sh}}^2\frac{1}{\xi^2}, \label{dim momentum}\\ 
    &n\gamma-n-1+\left(f_{\rm s}-\xi\right)\frac{d{}}{d{\xi}}\ln\left({\frac{h_{\rm s}}{{g_{\rm s}^{\gamma}}}}\right)=0, \label{dim energy eq}
\end{align}
while the jump conditions,
\begin{align}
    v\left(R\right) &= \frac{2}{\gamma+1}\left[1+\frac{\gamma-1}{2}\frac{v_{\rm a}}{V_{\rm s}}\right]V_{\rm s}, \\     
    \rho\left(R\right) &= \frac{\gamma+1}{\gamma-1}\rho_{\rm a}, \\ 
    p\left(R\right) &= \frac{2}{\gamma+1}\left[1-\frac{v_{\rm a}}{V_{\rm s}}\right]^2 \rho_{\rm a}V_{\rm s}^2,
\end{align}
give the values of the functions at the shock: 
\begin{align}
    f_{\rm s}\left(1\right) &= \frac{2}{\gamma+1}-\frac{3\eta_{\rm sh}}{2}\frac{\gamma-1}{\gamma+1}f_{\rm a}\left(\etash\right), \label{dim v jump}\\     
    g_{\rm s}\left(1\right) &= \frac{\gamma+1}{\gamma-1}g_{\rm a}\left(n, \etash\right), \label{dim dens jump}\\ 
    h_{\rm s}\left(1\right) &= \frac{2}{\gamma+1}\left[1+\frac{3\eta_{\rm sh}}{2}f_{\rm a}\left(\etash\right)\right]^2 g_{\rm a}\left(n, \etash\right). \label{dim p jump}
\end{align}
Note that $g_{\rm a}$ and $f_{\rm a}$ are the ambient density and velocity defined by Equations \eqref{amb dens} \& \eqref{amb v}. 

Equations \eqref{dim continuity} -- \eqref{dim energy eq} with the boundary conditions \eqref{dim v jump} -- \eqref{dim p jump} can be numerically integrated (for a given $n$ and $\gamma$) by specifying $\eta_{\rm sh}$, and --- similar to \citet{coughlin18} --- there is an ``eigenvalue,'' or a special value of $\eta_{\rm sh}$, that permits the smooth passage of the fluid variables through a sonic point within the post-shock flow and accretion onto the central object; these solutions are physically relevant to weak and failed explosions where the neutron star collapses to a black hole. The location of the sonic point (in terms of the self-similar variable $\xi$) can be derived by writing Equations \eqref{dim continuity} -- \eqref{dim energy eq} in the form
\begin{equation}
    \mathbf{M} \cdot \frac{\partial \boldsymbol{\omega}}{\partial \xi}=\mathbf{R},
\end{equation}
where 
\begin{equation}
    \mathbf{M}=\begin{pmatrix}
g_{\rm s} & f_{\rm s}-\xi & 0\\
f_{\rm s}-\xi & 0 & {g_{\rm s}}^{-1}\\
0 & -\gamma(f_{\rm s}-\xi){g_{\rm s}}^{-1} & (f_{\rm s}-\xi){h_{\rm s}}^{-1}\\
\end{pmatrix}
\end{equation}
is a matrix with eigenvalues that correspond to the characteristic speeds that information propagates within the post-shock fluid, $\boldsymbol{\omega}=(f_{\rm s},g_{\rm s},h_{\rm s})^{\mathbf{T}}$, and $\mathbf{R}$ is a vector that depends on $f_{\rm s}$, $g_{\rm s}$, $h_{\rm s}$, $\gamma$, $n$, and $\xi$. The sonic point occurs when $\mathbf{M}$ is singular, and therefore when
\begin{equation}
    \det(\mathbf{M})\propto (f_{\rm s}-\xi)^2g_{\rm s} - \gamma h_{\rm s} =0,
\end{equation}
which --- similar to Bondi flow --- is satisfied by a specific value of $\eta_{\rm sh}$ that maintains the regularity of the fluid variables across this point. We can straightforwardly solve for this value by shooting from the shock front. 

\section{Results}
\label{sec: results}
\begin{figure*} 
    \includegraphics[width=0.337\linewidth]{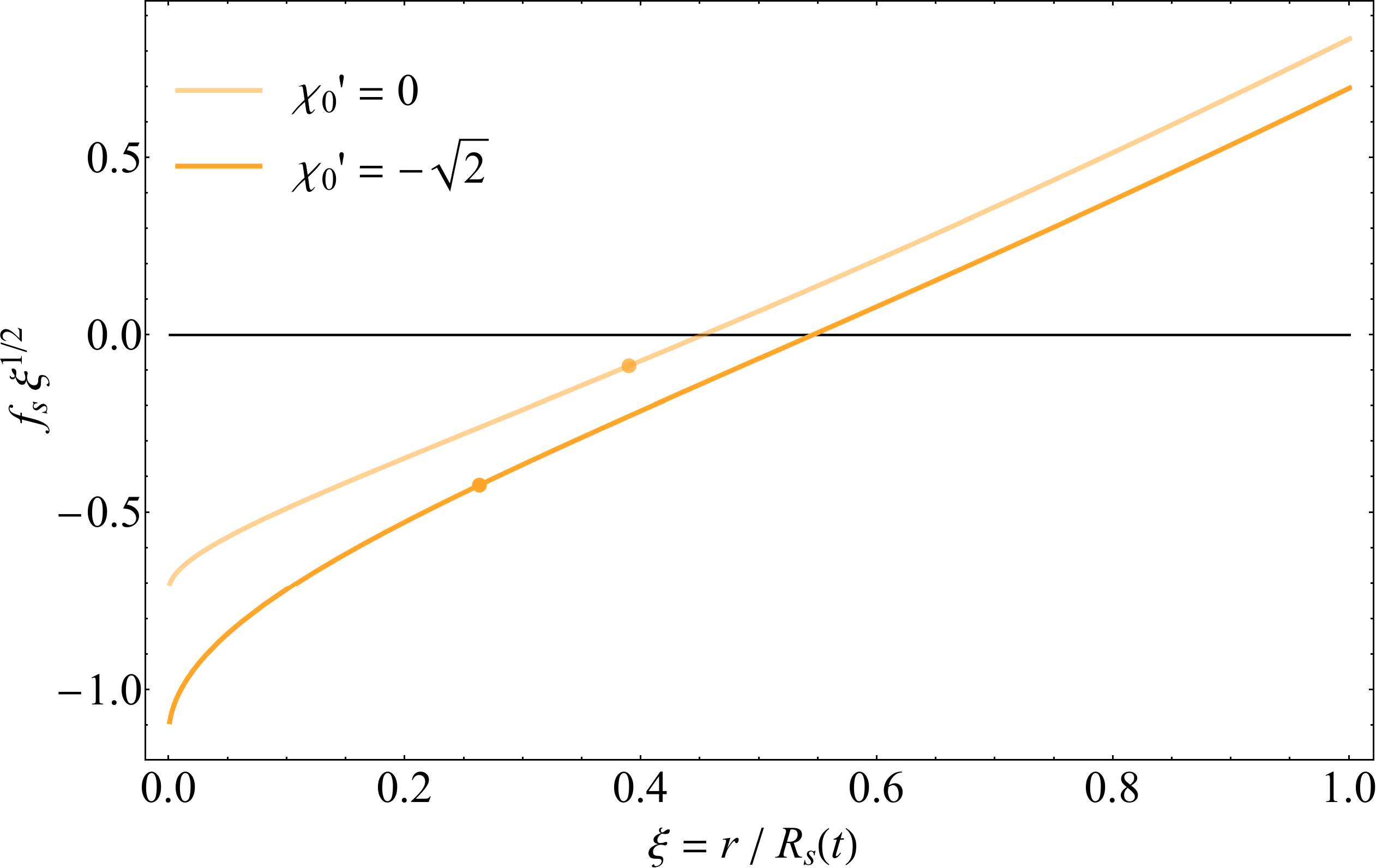}
    \includegraphics[width=0.3225\linewidth]{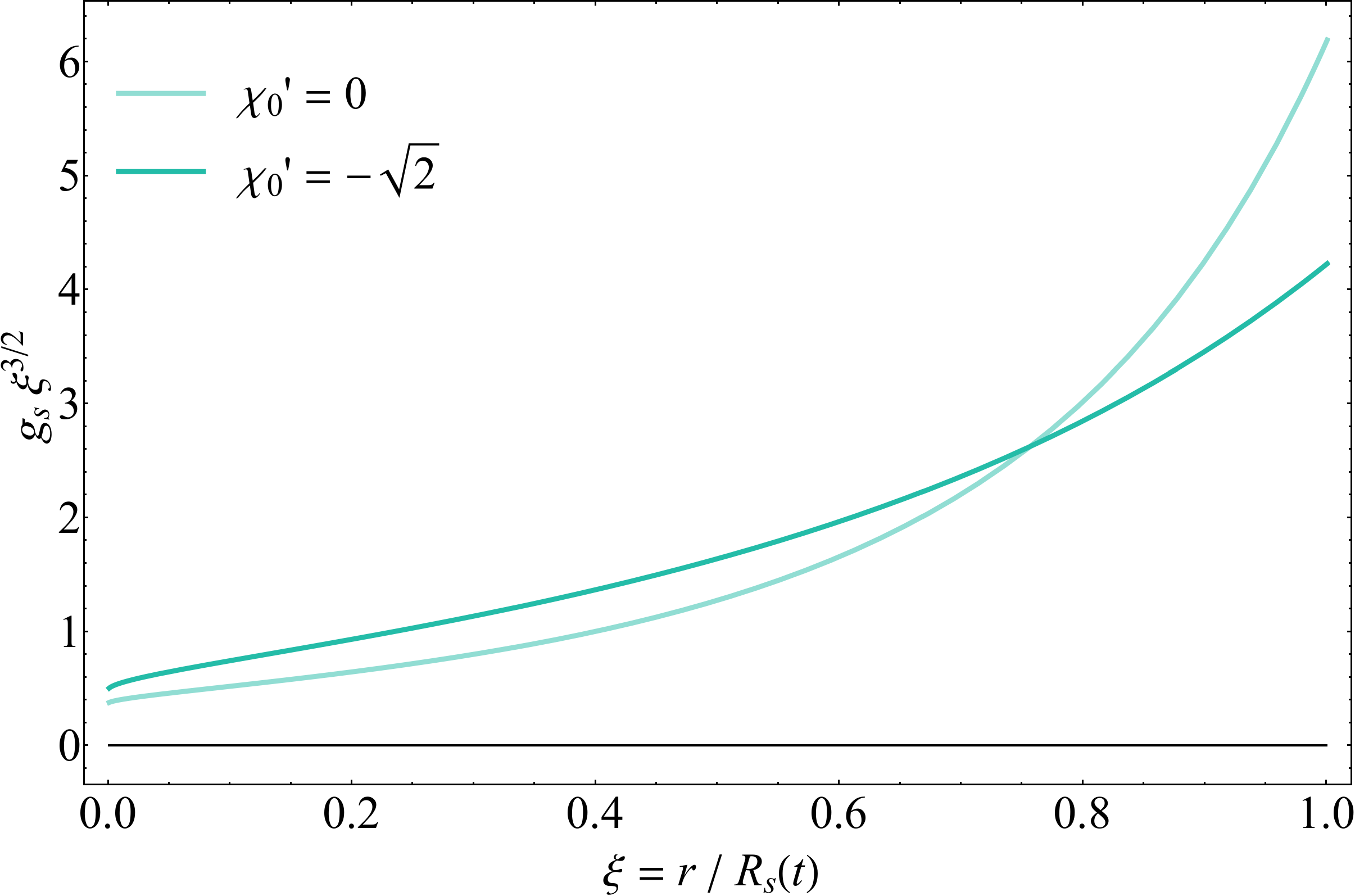}
    \includegraphics[width=0.3325\linewidth]{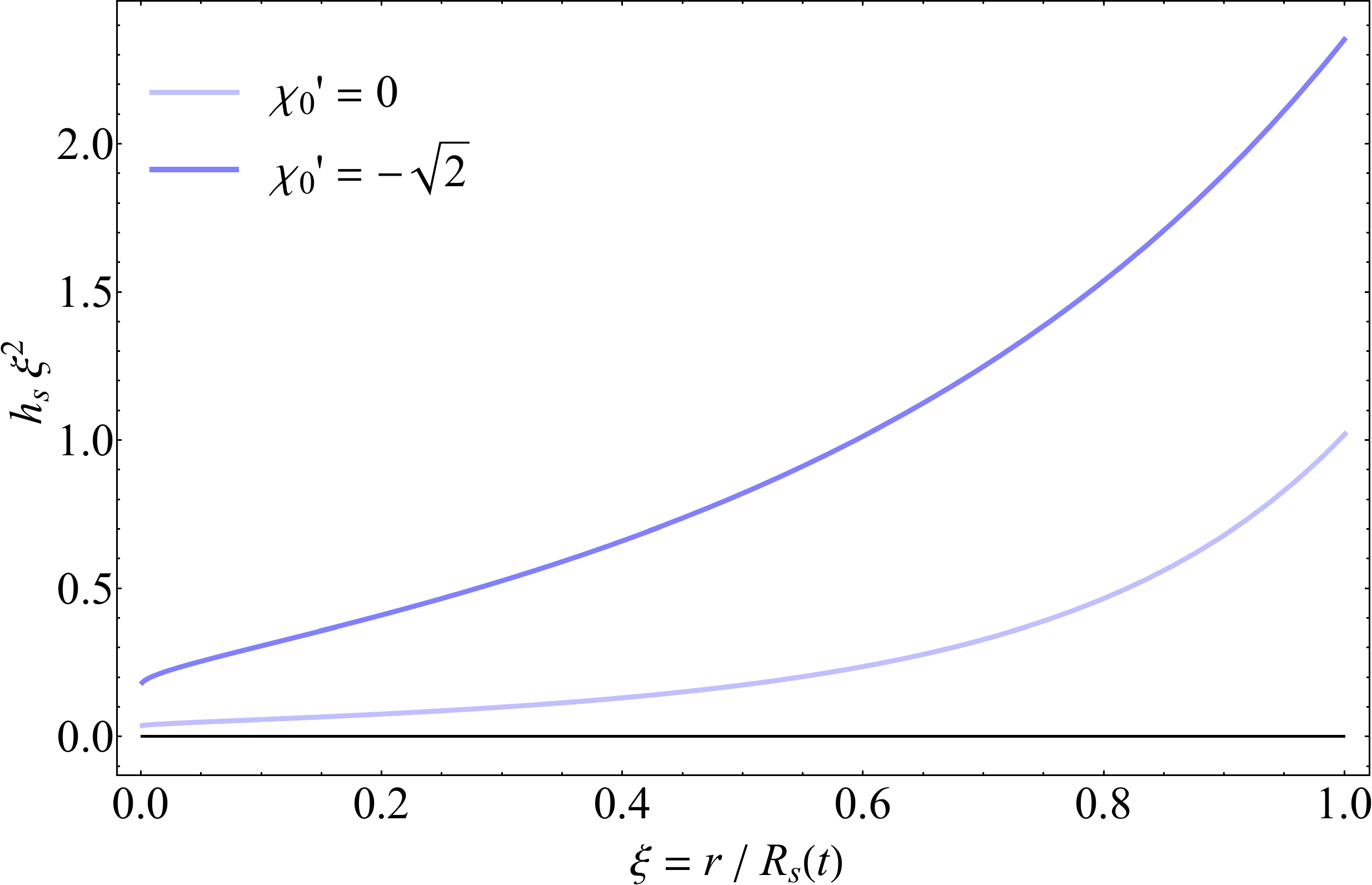}
\caption{The post-shock self-similar velocity (left), density (middle), and pressure (right), as functions of the self-similar variable $\xi = r / R_{\rm s}\left(t\right)$, normalized by their scalings as $\xi\rightarrow0$, which are ${\xi}^{-1/2}$, ${\xi}^{-3/2}$, and ${\xi}^{-2}$, respectively, for $n=2.5$ and $\gamma=4/3$. The dark and light curves indicate the solutions for the ambient gas starting out in freefall ($\chi_0'=-\sqrt{2}$) and at rest ($\chi_0'=0$). The filled points on the velocity ($f_{\rm s}$) curves indicate the  sonic point.} \label{fig:fgh43}
\end{figure*}
Figure \ref{fig:fgh43} shows the post-shock fluid velocity, density, and pressure profiles for the limiting cases of $\chi'_0=0$ (light curves) and $\chi'_0=-\sqrt{2}$ (dark curves), with an ambient power-law index of $n=2.5$ and adiabatic index of $\gamma=4/3$; the latter is valid while the shock is still deep in the interior and the gas is radiation pressure dominated and/or the electrons are relativistic and contribute substantially to the pressure support. Each curve is normalized by its scaling near the origin, with the velocity and density approaching freefall (i.e., $f_{\rm s} \propto {\xi}^{-1/2}$ and $g_{\rm s} \propto {\xi}^{-3/2}$) and the pressure conforming to the adiabatic scaling (i.e., $h_{\rm s}\propto {g_{\rm s}}^{\gamma}$). The left panel shows that the fluid velocity near the shock front is positive, such that the gas initially moves outward after being hit by the shock, but ultimately becomes negative with gas accreting onto the central object near the origin. Each solution also passes through the sonic point, which is indicated by a filled point on each $f_{\rm s}$ curve. While the sonic point for the rest case resides near the stagnation ($v=0$) point, the sonic point for the freefall case is both at a smaller radius than the rest case and at a more negative velocity. This can be attributed to the post-shock pressure (right panel) being larger for the freefall solution, whereas the post-shock densities (middle panel) remain comparable, thereby requiring a larger (in magnitude) fluid velocity to reach supersonic speeds. Additionally, while solutions with different values of $\gamma$ can be found, they are both qualitatively and quantitatively similar to the results shown in Figure \ref{fig:fgh43}. We therefore do not include them here and our fiducial case for the remainder of the paper will be $\gamma=4/3$ unless stated otherwise.

\begin{figure*} 
    \includegraphics[width=0.495\linewidth]{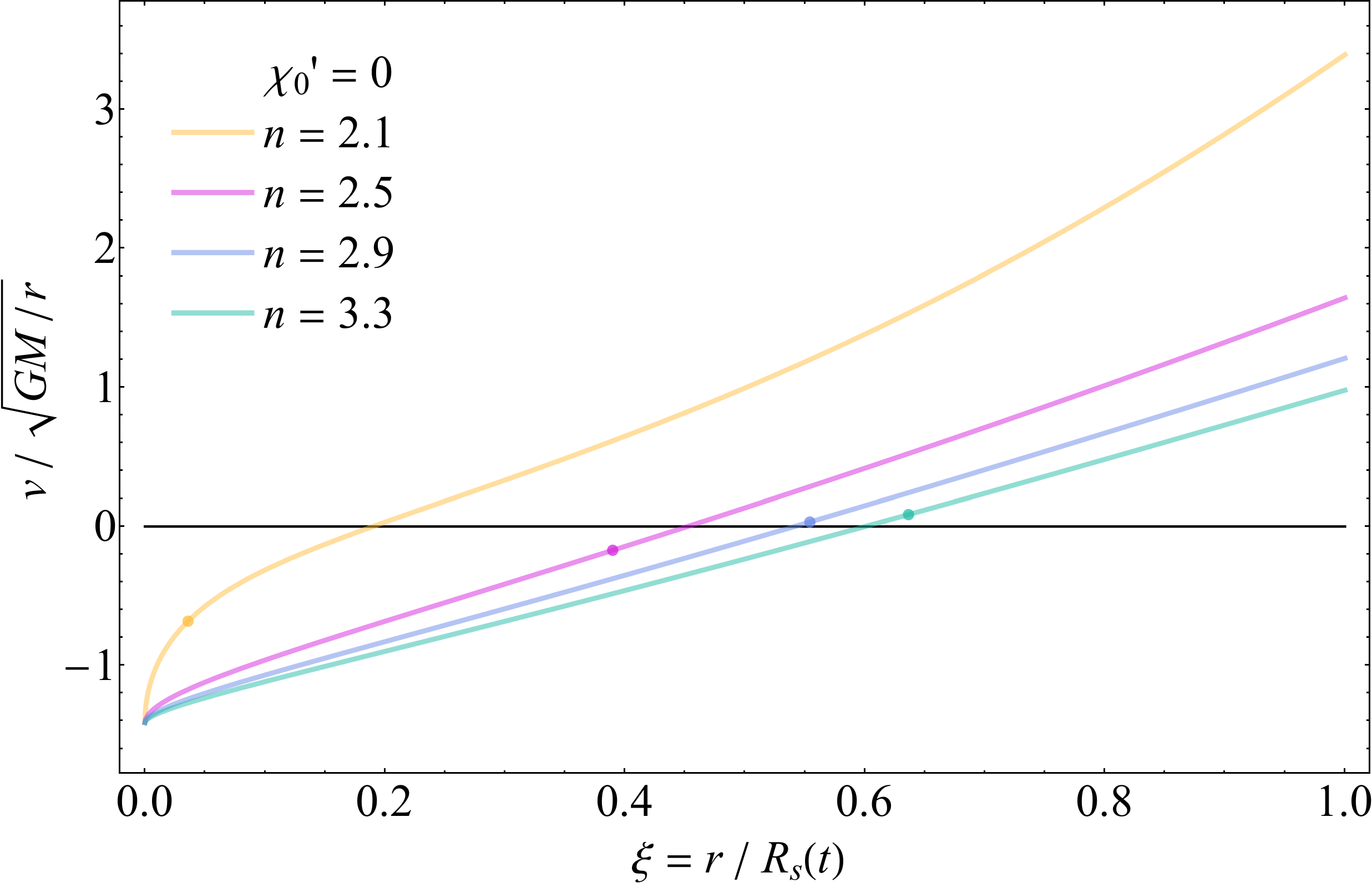}
    \includegraphics[width=0.4975\linewidth]{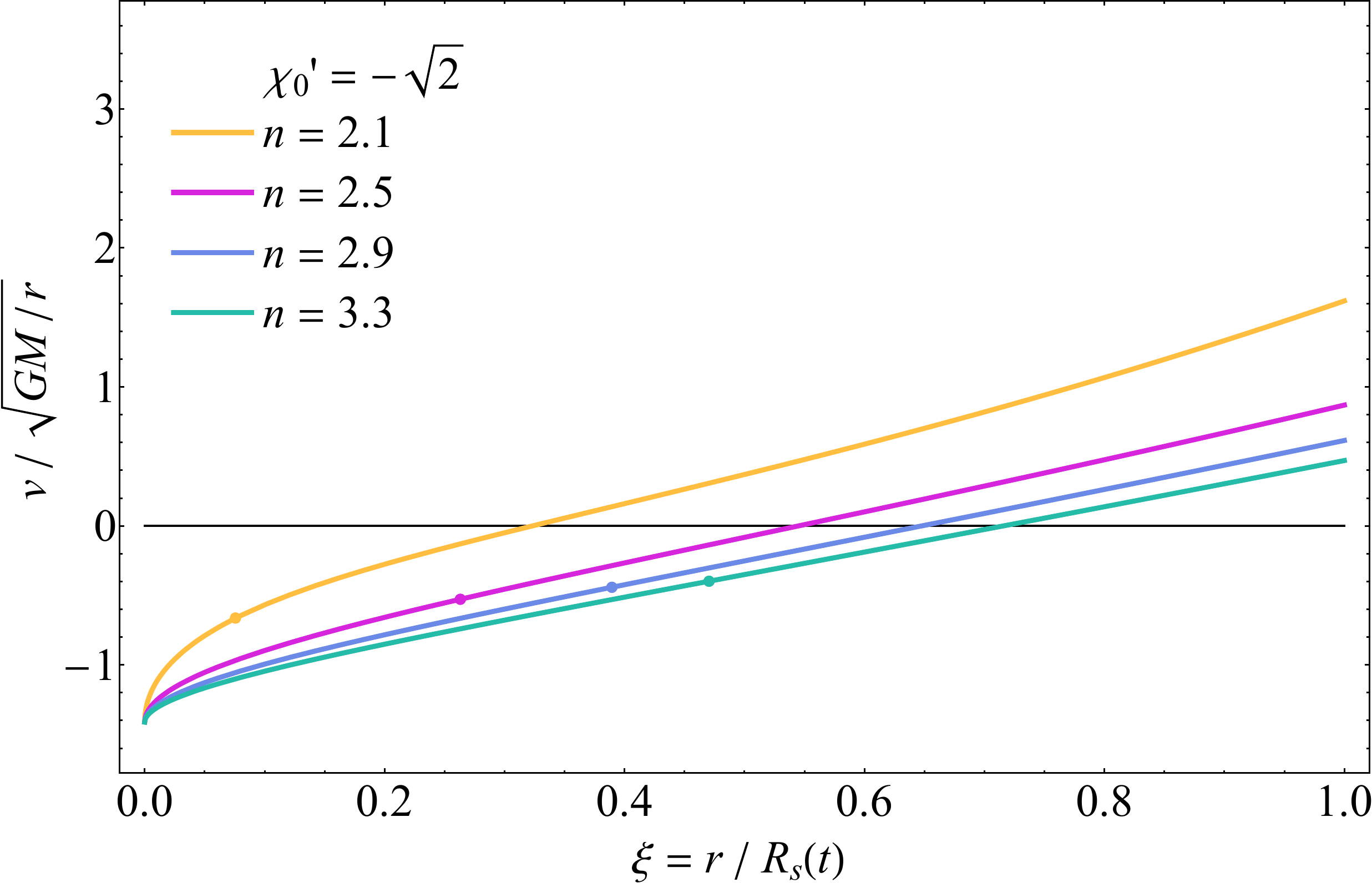}
\caption{The post-shock fluid velocity normalized by $\sqrt{GM/r}$ with $\gamma=4/3$ and four different values of the power-law index $n$ when the ambient medium is asymptotically at rest (left panel) and in freefall (right panel). The fluid velocity near the shock is positive and larger (smaller) for shallower (steeper) power-law indices. Additionally, the location of the sonic point in the post-shock flow --- indicated by the filled in points on each curve --- shifts toward smaller radii for smaller values of $n$. All solutions approach pressureless freefall at small radii.} \label{fig:f_ns}
\end{figure*}

Figure \ref{fig:f_ns} shows the post-shock fluid velocity (normalized by the Keplerian velocity) for four different power-law indices, $\gamma=4/3$, $\chi_0'=0$ (left panel), and $\chi_0'=-\sqrt{2}$ (right panel). Each curve shows the same qualitative behavior --- positive velocity near the shock front, a stagnation point, passage through a sonic point, and freefall near the origin --- but the velocity is noticeably and everywhere larger as $n$ decreases. It is also interesting to note that, for power-law indices greater than $n\sim2.8$, the sonic point is located in the positive region of the post-shock flow for the at-rest case, whereas the sonic points for the freefall case are always located within the negative velocity region. 

\begin{figure*}
    \includegraphics[width=0.497\textwidth]{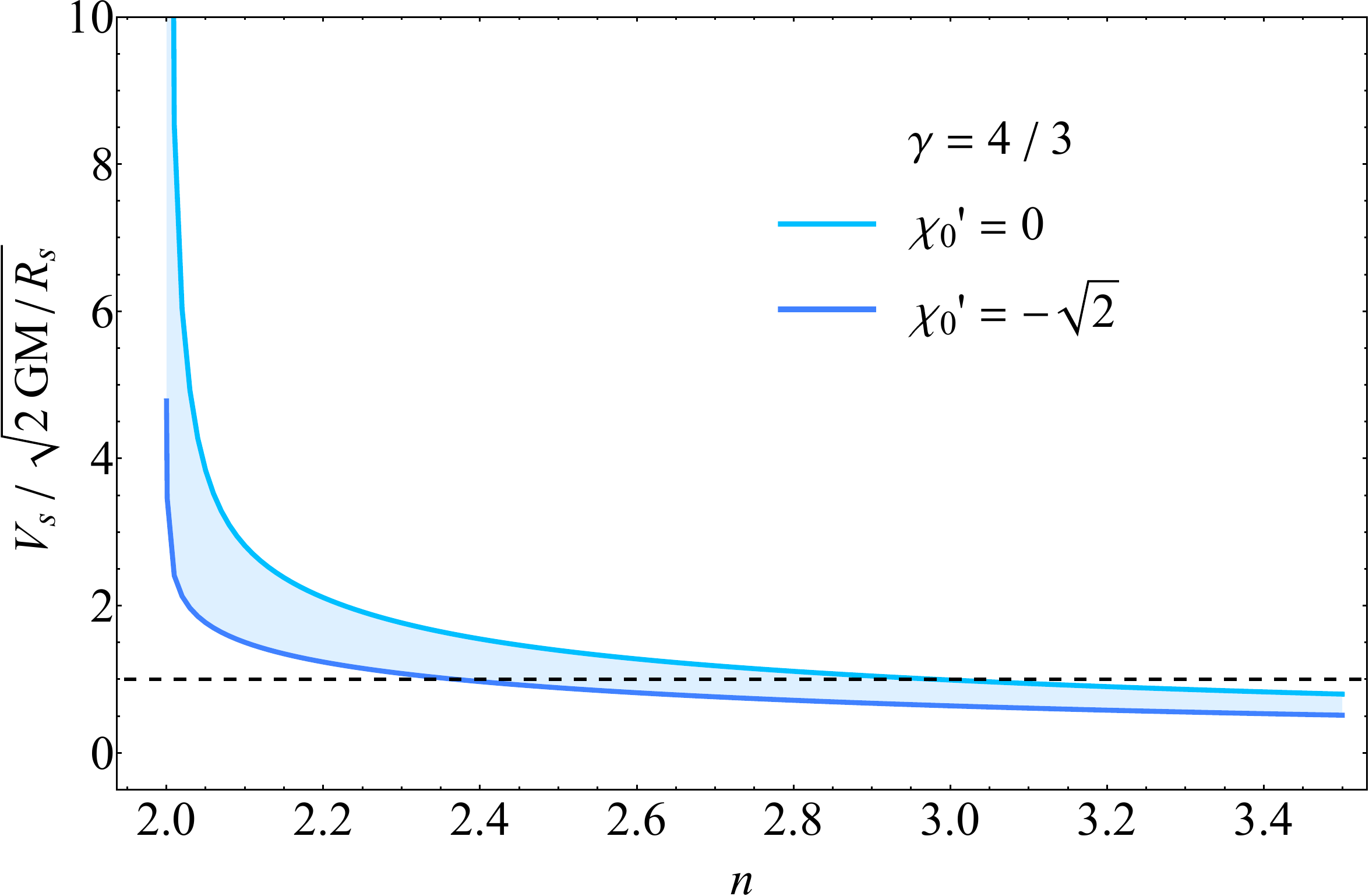}
    \includegraphics[width=0.4925\textwidth]{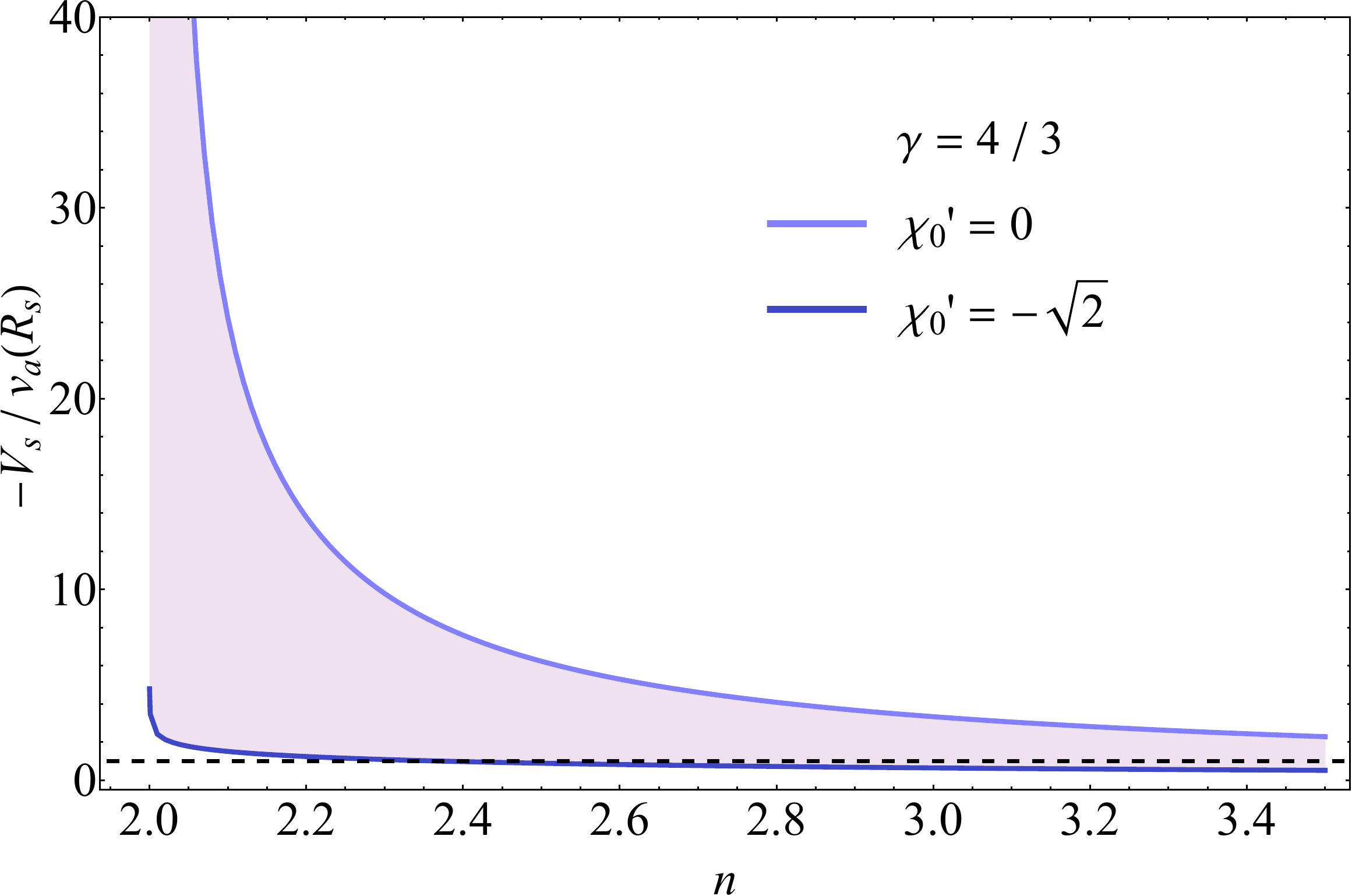}
    \caption{Left: the critical shock speed normalized by the escape speed as a function of the power-law index $n$. As $n$ increases $V_{\rm s}$ decreases and is eventually below the Keplerian escape velocity, shown by the black-dashed line. The light and dark blue curves represent the cases where the ambient medium is initially static ($\chi_0'=0$) and initially in freefall ($\chi_0'=-\sqrt{2}$), respectively, while the shaded region represents cases where the initial velocity of the ambient medium is in between these two limits. Right: the negative of the self-similar shock velocity normalized by the ambient velocity --- given by Equation \eqref{amb v} --- at the location of the shock as a function of the power-law index $n$, with the limiting cases of the ambient medium at rest and in freefall shown by the light and dark purple curves, respectively. The shock speed (as well as its ratio to the ambient speed) diverges as $n\rightarrow 2$.}
    \label{fig:Vcrit_n}
\end{figure*}
\begin{figure}
    \includegraphics[width=0.47\textwidth]{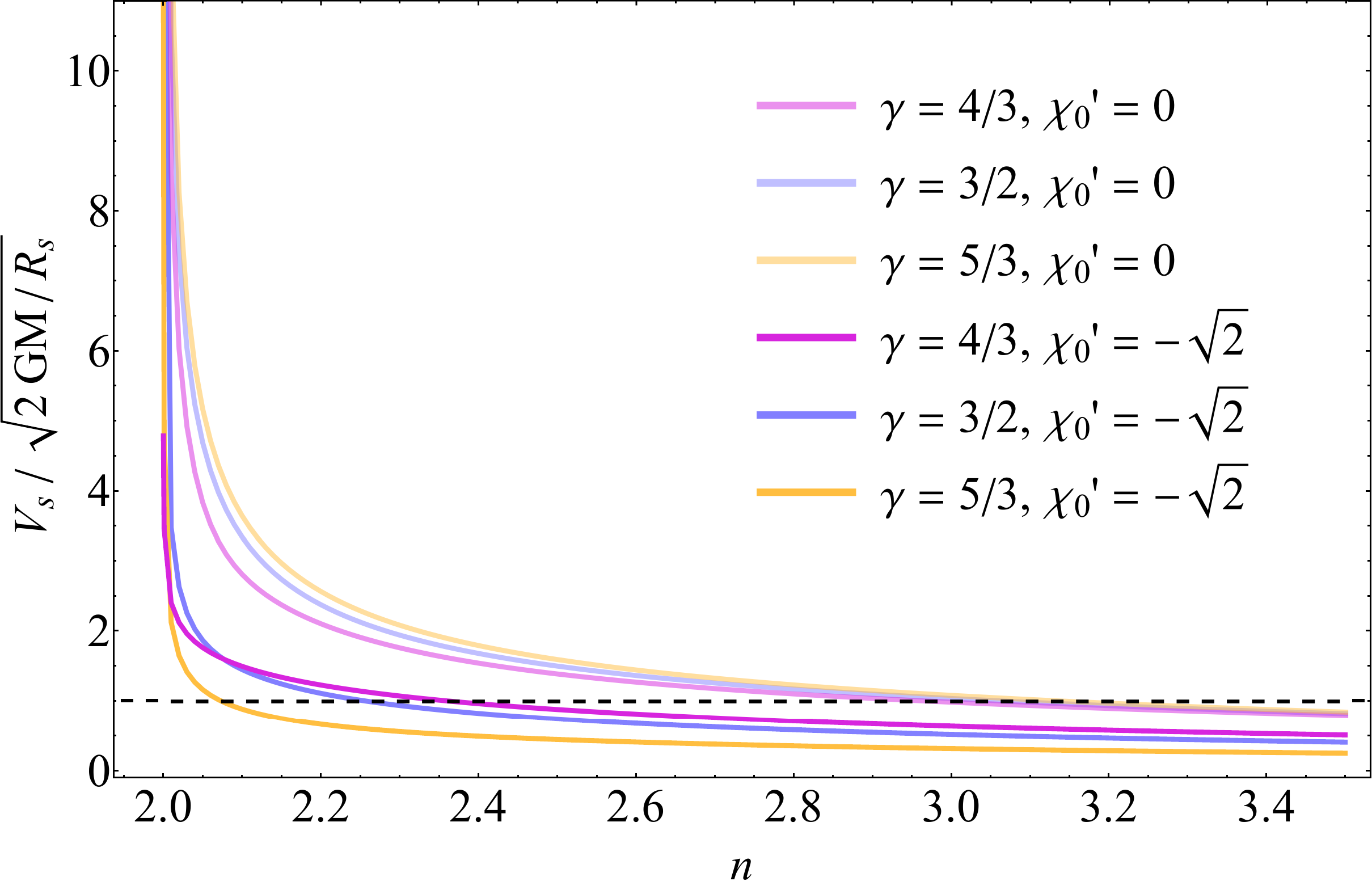}
    \caption{Same as the left panel of Figure \ref{fig:Vcrit_n}, but for different adiabatic indices as well. All curves exhibit the same qualitative trend, and the value of the adiabatic index does not strongly affect the critical shock speed.}
    \label{fig:Vcrit_n_gamma}
\end{figure}

In the left panel Figure \ref{fig:Vcrit_n} we show the self-similar shock velocity --- as predicted by Equation \eqref{RV shock} --- normalized by the local escape velocity as a function of $n$ for $\gamma=4/3$. The light and dark blue curves show the solutions for when the ambient fluid starts out at rest and freefall, respectively; the shaded region is occupied for ambient velocities in between the rest and freefall cases, i.e., the instantaneous-freefall and at-rest cases provide lower and upper limits to the shock speed, respectively. The black, dashed line shows when the ratio is unity, and therefore indicates the division between super and sub-Keplerian shock velocities. Shallower density profiles (i.e., smaller values of $n$) require a larger shock velocity, which is reasonable if these solutions represent the fluid analog of marginally bound orbits, as ambient media with larger $n$ have smaller gravitational binding energies, rendering the shock more readily capable of escaping. The right panel of Figure \ref{fig:Vcrit_n} shows the negative ratio of the self-similar shock velocity to the ambient velocity at the location of the shock front. Differences between the two solutions are more pronounced in this plot compared to the left panel, as the velocity of the shock for the at-rest case is larger than the freefall solution, while the ambient velocity is smaller (larger) in magnitude for the rest (freefall) solution. For all ambient velocities, we find that the shock velocity diverges in the limit that $n\rightarrow2$; we return to this point in Section \ref{sec: discussion}. Figure \ref{fig:Vcrit_n_gamma} is analogous to Figure \ref{fig:Vcrit_n}, but varies $\gamma$ according to the legend, illustrating that the general trend of $V_{\rm s}(n)$ is unaffected by the adiabatic index. 

\section{Numerical Simulations}
\label{sec: simulations}
To verify that our self-similar solutions exactly solve the fluid equations, we perform numerical simulations with the finite-volume hydrodynamics code \textsc{flash} (V4.7; \citealt{fryxell00}). We use a uniform spherical grid with $10^5$ cells, an inner and outer radius of $r_{\rm in}=0.1$ and $r_{\rm out}=100$, and we adopt normalized units such that $2GM=1$. We use outflow (zero-gradient) and reflecting boundary conditions at the inner and outer boundaries, respectively; the latter generates non-physical effects near the outer boundary and at late times, but we are only interested in the accuracy of the self-similar solutions at times well before this, i.e., these effects do not impact our results. We initialize the simulations with the self-similar solutions for $n = 2.5$, $\gamma = 4/3$, $\chi'_0 = -\sqrt{2}$, and when the shock is at $R_{\rm s} = R_{\rm i} = 1$; the ambient pressure is floored at $10^{-10}$. Time is thus measured in units of the freefall time from the initial radius of the shock, 
\begin{equation}
\tau_{\rm ff} = \frac{R_{\rm i}^{3/2}}{\sqrt{2GM}}.
\end{equation}

\begin{figure}
    \includegraphics[width=0.47\textwidth]{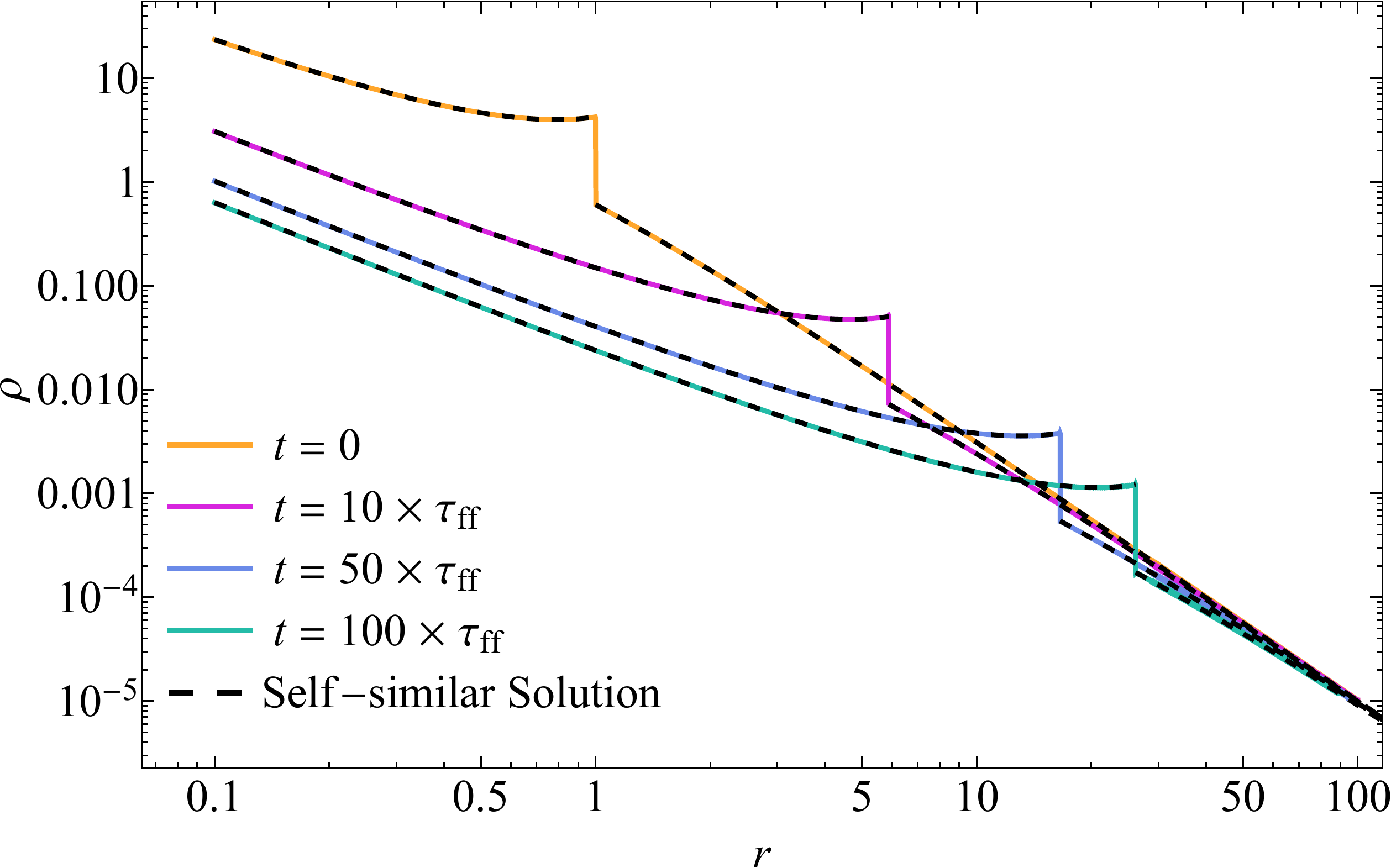}
    \caption{The density profiles from the self-similar solution (dashed) and a numerical simulation ran with \textsc{flash} (solid) at different times.}
    \label{fig: num dens comp}
\end{figure}

\begin{figure}
    \includegraphics[width=0.47\textwidth]{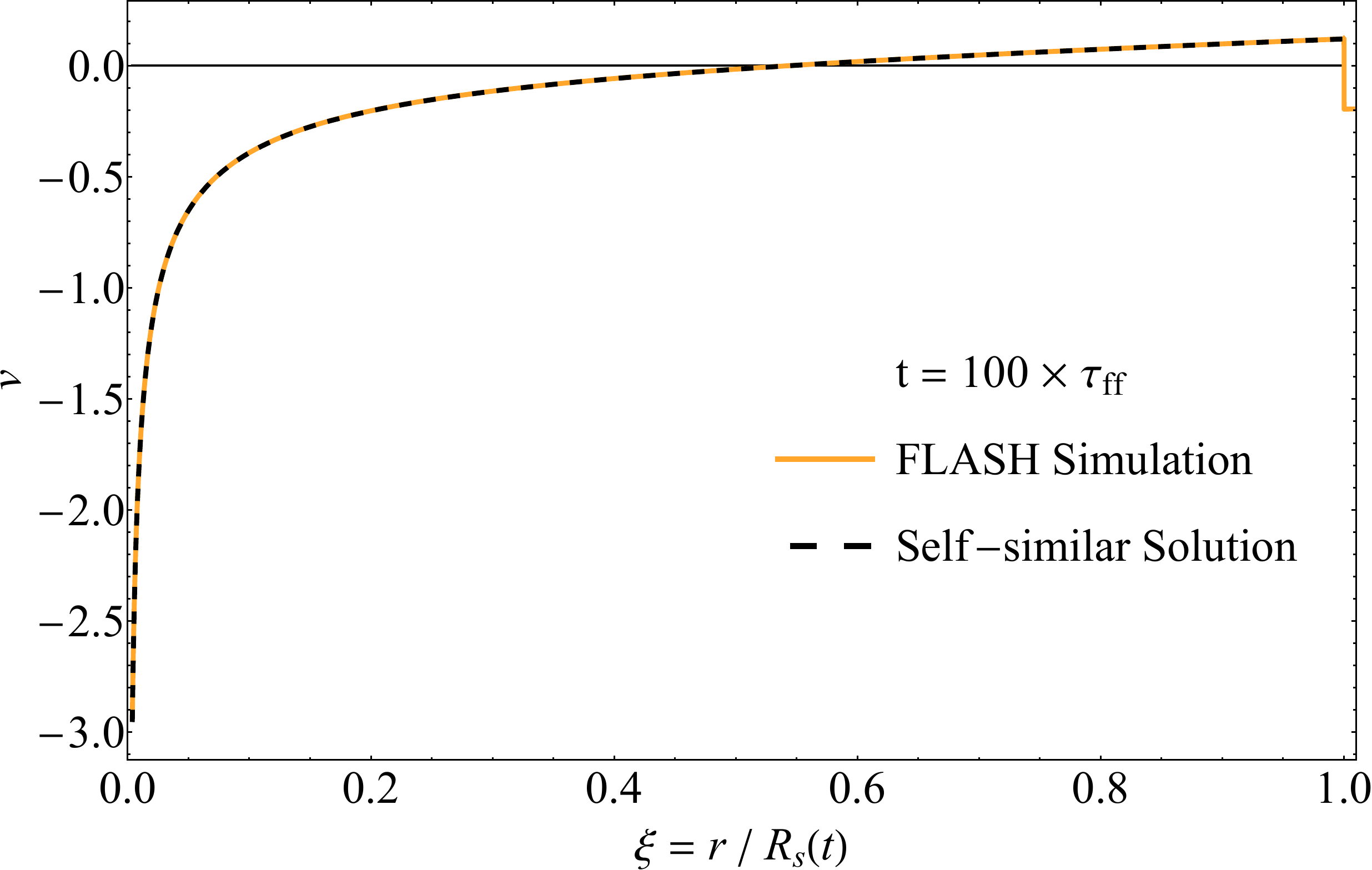}
    \caption{The fluid velocity as a function of the self-similar radius (i.e., $r/R_{\rm s}$) at a time of $t=100\times \tau_{\rm ff}$. The numerical solution obtained from \textsc{flash} is shown in yellow, while the analytical prediction is given by the black-dashed curve.}
    \label{fig: num v comp}
\end{figure}

Figure \ref{fig: num dens comp} compares the numerically obtained (solid) and analytically predicted (dashed) density at the times in the legend, illustrating that the two are in excellent agreement after many dynamical times, thus effectively verifying that the analytical solution exactly satisfies the fluid equations. Figure \ref{fig: num v comp} compares the numerically obtained post-shock velocity and the self-similar prediction as a function of radius at $t=100\times\tau_{\rm ff}$, again showing near-exact agreement. Despite this agreement, we expect the analytical solutions to represent fluid-dynamical analogs of marginally bound orbits, and hence the numerical solutions should asymptotically deviate from them owing to small numerical errors; we return to this point in Section \ref{sec:stability} below.


\section{Discussion}
\label{sec: discussion}
\subsection{Energy}
As the shockwave propagates outward it sweeps up (negative) binding energy from the ambient medium, while the black hole removes energy from the system through accretion. Therefore, unless the rate at which the black hole accretes energy exactly balances the rate at which the shock entrains binding energy, we expect the energy within the post-shock flow to be time-dependent. Specifically, the energy of a region of the post-shock fluid is given by 
\begin{equation}
    \label{E int}
    E = 4\pi\int_{R_0\left(t\right)}^{R_{\rm s}\left(t\right)} \left[\frac{1}{2}v^2+\frac{1}{\gamma-1}\frac{p}{\rho}-\frac{GM}{r}\right]\rho r^2dr,
\end{equation}
where $R_0\left(t\right)$ is a radius that satisfies $0\leq R_0\left(t\right) < R_{\rm s}\left(t\right)$. This integral can be evaluated over the entirety of the shocked gas (i.e., $R_0=0$). However, we expect solutions that have a shock speed slightly larger than the critical one identified here to eventually transition to the strong/ST regime, for which only the material with positive velocities will eventually contribute to the conserved energy, with everything interior to the stagnation radius accreting onto the compact object\footnote{As for the solutions found in \citet{coughlin18}, self-similarity requires that every fluid element eventually falls back to the origin, and hence by truncating the integral at the stagnation radius we are effectively providing an upper limit to the final energy.}. We therefore define $R_0\left(t\right)$ as the radius where $v\left(R_0\left[t\right]\right)=0$, for which 
\begin{equation}
    \label{energy}
    E = 4\pi\rho_{\rm i}{r_{\rm i}}^3\frac{GM}{r_{\rm i}}\left(\frac{R_{\rm s}}{r_{\rm i}}\right)^{2-n}E_{\rm s} \propto t^{\frac{4}{3}-\frac{2n}{3}},
\end{equation}
where in the last line we defined
\begin{equation}
    \label{Es}
    E_{\rm s}\equiv\frac{4}{9{\etash}^2}\int_{\xi_0}^1 \left[\frac{1}{2}{f_{\rm s}}^2+\frac{1}{\gamma-1}\frac{h_{\rm s}}{g_{\rm s}}-\frac{9{\etash}^2}{4\xi}\right]g_{\rm s}\xi^2d\xi.
\end{equation}
Here $\xi_0$ is the self-similar stagnation radius, i.e., $f_{\rm s}\left(\xi_0\right)=0$. For values of $n>2$ --- the range of power-law indices analyzed here --- the post-shock energy is a decreasing function of time, and approaches zero in the limit that the shock propagates to infinity. 

\begin{figure}
    \includegraphics[width=0.47\textwidth]{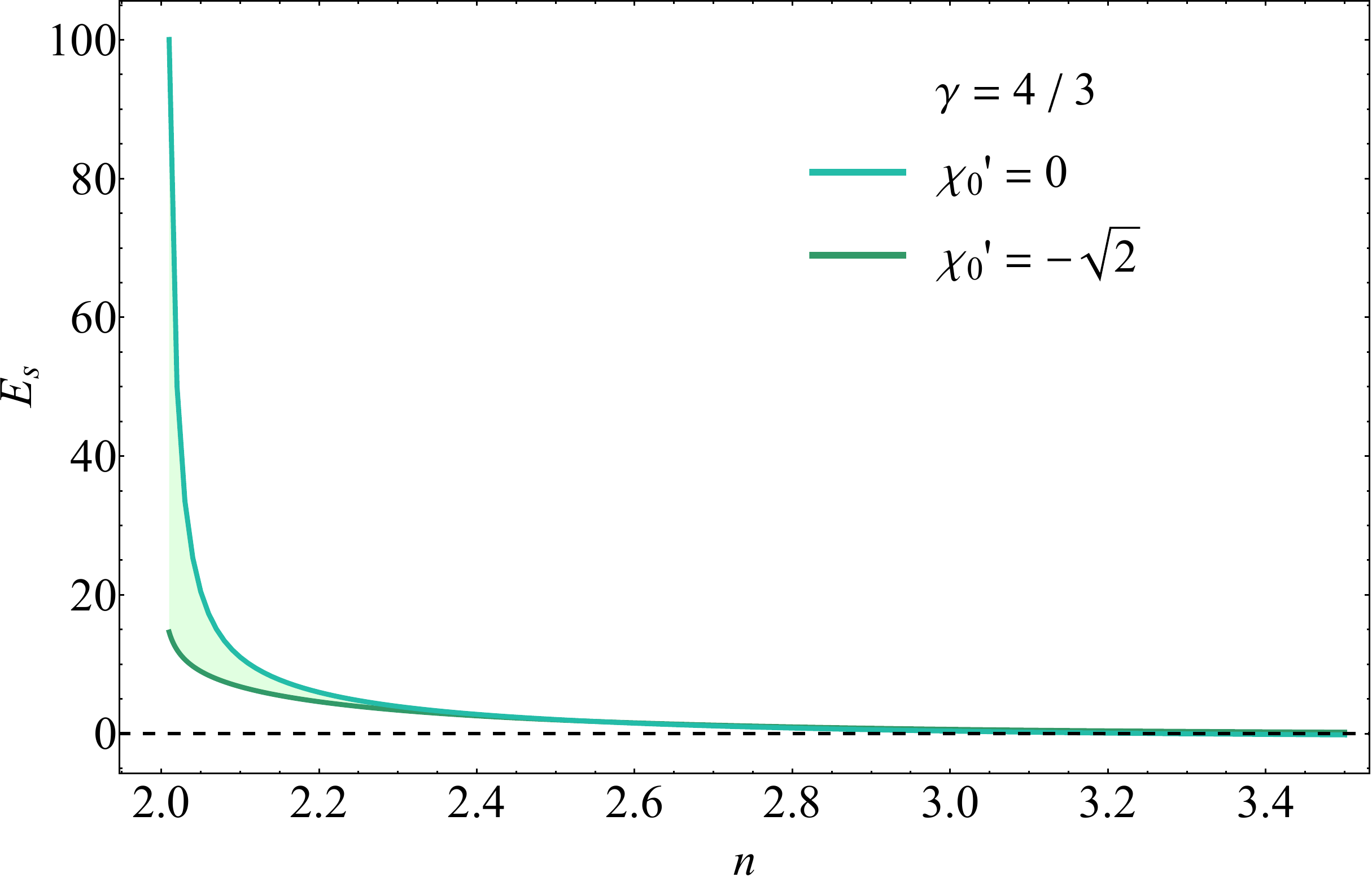}
    \caption{The energy integral given by Equation \eqref{Es} as a function of  $n$ and $\gamma=4/3$. The light and dark curves show the solutions for an initially at-rest ($\chi_0'=0$) and freefalling ($\chi_0'=-\sqrt{2}$) ambient medium, respectively, and the shaded region represents solutions with an ambient speed between these cases. As $n\rightarrow2$, the energy diverges to infinity, whereas it is at most of the order unity for most values of $n$. For $n\gtrsim 2.23$ the solutions become nearly indistinguishable, with the at-rest solution crossing over the freefall solution at $n\sim2.55$ before becoming negative for $n\gtrsim3.27$.}
    \label{fig: Es}
\end{figure}
We plot the results of this integration for $\gamma=4/3$ and a range of $\chi_0'$ values as a function of $n$ in Figure \ref{fig: Es}. Similar to what is shown in Figures \ref{fig:Vcrit_n} \& \ref{fig:Vcrit_n_gamma}, noticeable differences between the solutions arise in the limit that $n\rightarrow2$, and the energy diverges in this limit. For values of $n\gtrsim2.23$, the value of $E_{\rm s}$ is $\lesssim 1$, and the different ambient solutions are nearly indistinguishable, with differences between the two solutions being at most of the order $10^{-1}$. It is worth noting, however, that the solutions cross at $n~2.55$, and the at-rest ($\chi_0'=0$) solution is negative for $n\gtrsim3.27$.

\subsection{Solutions with n=2}
\begin{figure}
    \includegraphics[width=0.47\textwidth]{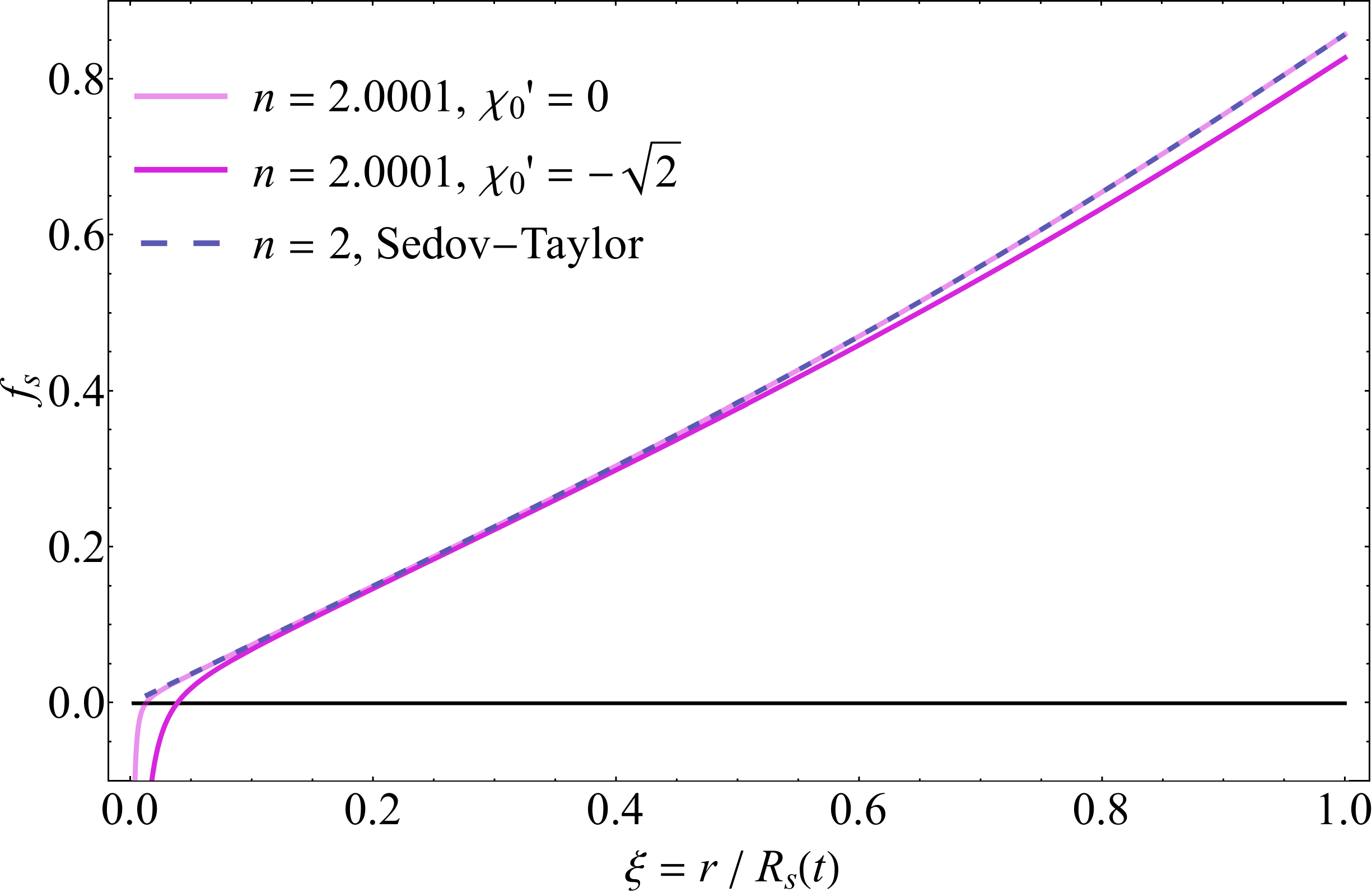}
    \caption{The self-similar post-shock velocity with $n=2.0001$ and $\gamma=4/3$ for the initially at-rest (light) and infalling (dark) solutions compared to the $n=2$ and $\gamma=4/3$ Sedov-Taylor solution (dashed). This demonstrates that our solutions approach the Sedov-Taylor solution in the limit that $n\rightarrow2$.}
    \label{fig: ST Comp}
\end{figure}
Figures \ref{fig:Vcrit_n} \& \ref{fig: Es} show that the self-similar shock velocity and energy diverge in the limit that $n\rightarrow2$; the sonic point also approaches the origin in this same limit (see Figure \ref{fig:f_ns}). This behavior can be understood by examining the gravitational potential energy of the ambient medium,
\begin{align}
    \label{Egrav}
    E_{\rm grav}=-\int_{0}^{r}\frac{GM}{r}\rho r^2 dr \propto\frac{1}{2-n}r^{2-n},
\end{align}
 which diverges for $n \le 2$. The shock would thus need infinite energy in order to propagate to infinity. In addition, the post-shock energy given by Equation \eqref{energy} is time-independent when $n=2$, and is therefore a conserved quantity. The only self-similar solution that can exist for $n = 2$ is therefore the ST solution. 

To demonstrate this directly, Figure \ref{fig: ST Comp} compares the self-similar post-shock velocity from our solution with $n=2.0001$ and $\gamma=4/3$ for both the at-rest and freefall cases (indicated by the light and dark curves, respectively) to the $n=2$ ST solution. For this specific combination of $n$ and $\gamma$, the sonic point is located at $\xi\simeq 0.001$ ($\simeq0.002$) for the rest (freefall) solution and the corresponding critical shock velocity is $V_{\rm s} / \sqrt{2GM/R_{\rm s}} \simeq 86$ ($\simeq4.8$). The close resemblance between our solution and the ST solution further solidifies the notion that our solutions approach the ST solution as $n\rightarrow 2$. 

The asymptotic behavior of $V_{\rm s} \rightarrow \infty$ as $n \rightarrow 2$,the divergence of the gravitational energy of the ambient medium for $n \le 2$, and the numerical verification of our solutions in Section \ref{sec: simulations} suggest that there are no self-similar solutions of the type described here --- those with a finite ratio of the shock speed to the freefall speed that pass through a critical point and accrete onto the central object --- for $n \le 2$ and general $\gamma$. When the ambient gas has an adiabatic index of $\gamma=4/3$ and a power-law index of $n=2$, the ambient medium has zero binding energy, and an energy-conserving solution that does not contain a sonic point can be found \citep{chevalier89}. 

\citet{Yalinewich21} relied on similar arguments --- namely that a point-mass gravitational field does not introduce a new velocity scaling with radius when $n=2$ --- to derive solutions that describe a shock propagating into an infalling, initially $\rho\propto r^{-2}$ ambient medium. Their solutions therefore appear to be similar to ours, but there are subtleties associated with their analysis. First, they considered adiabatic indices other than $\gamma=4/3$, meaning that the shock entrains negative energy for $\gamma > 4/3$ and positive energy for $\gamma<4/3$. The consequence of this is that their solutions must either accrete and remove energy from the system, or \emph{supply} energy to balance the rate that energy is being swept up, causing the solutions to end in a contact discontinuity (see the analogous discussion in Section II of \citealt{chevalier89}). They ultimately find that there is a critical energy below which the shock stalls, which they use to place constraints on the minimum energy for a successful supernova explosion. This claim is problematic considering, as discussed above, the binding energy of the ambient medium diverges when $n=2$. Consequently, the shock would need an infinite amount of energy to propagate to infinity, and therefore their solutions cannot be self-similar. In their Section 3, they discuss the divergence of the ambient energy, but remark that it can approximately be treated as a constant due to logarithmic divergence. Here, however, we show that this cannot be the case, and the divergence of the ambient binding energy consequently requires the shock to have an infinite velocity. 

\subsection{Stability}
\label{sec:stability}
Our solutions predict that the shock speed is proportional to the Keplerian velocity, and the total energy of the shocked fluid approaches zero as the shock recedes to infinitely large distances from the compact object. We therefore interpret them as fluid-dynamical analogs of marginally bound (zero energy) Keplerian orbits, such that the shock velocity shown in Figure \ref{fig:Vcrit_n} (and Figure \ref{fig:Vcrit_n_gamma}) is the necessary and \emph{minimum} velocity that a shock must attain in order to transition to the strong/ST regime. Additionally, any small change in the speed of the shock should lead to its stalling in the gravitational field of the compact object or transitioning to the Sedov-Taylor blastwave, i.e., these solutions should be unstable to radial perturbations. 

\begin{figure}
    \includegraphics[width=0.47\textwidth]{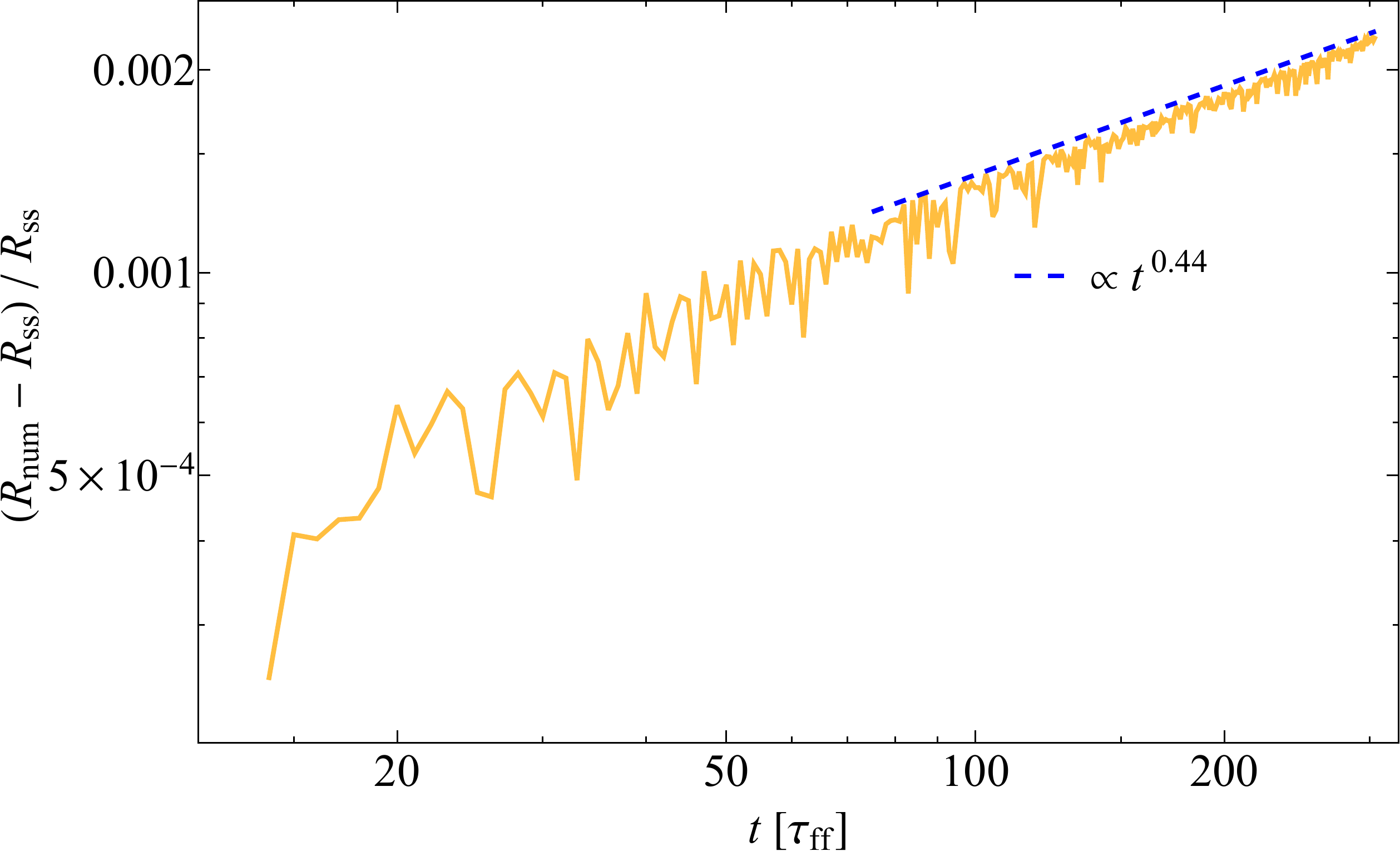}
    \caption{The relative difference between the numerically obtained ($R_{\rm num}$) and the analytically predicted ($R_{\rm ss}$) shock position as a function of time. The blue-dashed curve shows the approximate rate at which the two diverge.}
    \label{fig:R_diff Plot}
\end{figure}

In support of this conclusion, Figure \ref{fig:R_diff Plot} shows the relative difference between the numerical and analytical shock position as a function of time for the numerical simulation discussed in Section \ref{sec: simulations}. Even though Figures \ref{fig: num dens comp} and \ref{fig: num v comp} show excellent agreement between the two solutions, and thereby indirectly demonstrate that the solutions found here exactly solve the fluid equations, Figure \ref{fig:R_diff Plot} illustrates that the numerically obtained solution is slowly deviating from the self-similar solution. The rate at which the two solutions deviate from one another is approximately a power-law in time, i.e., this supports the claim that these solutions are indeed unstable, and the power-law growth of the instability is expected (e.g., \citealt{Ryu87, goodman90, sari00, coughlin19}). Here these deviations arise from numerical error, but in physical systems these can be introduced naturally through slight differences in, e.g., the ambient profile or the initial energy of the explosion. While Figure \ref{fig:R_diff Plot} provides evidence that these solutions are unstable, a more formal stability analysis is required to assess the wide range in parameter space that these solutions encompass (e.g., power-law index, initial ambient velocity), and can be accomplished using the same method that is outlined in Section 5 of \cite{coughlin23}. 

Furthermore, while we restricted our analysis to spherical symmetry, observational (e.g., \citealt{Wang08, Nagao19, Harrison75, Hobbs05}) and theoretical (e.g., \citealt{Burrows96, Khokhlov99, Foglizzo02, Blondin03, Scheck06, Nordhaus10, Wongwathanarat13, Janka17}) work shows that the core-collapse process is aspherical, and deviations from spherical symmetry can (through, for example, the convective turbulence in the post-shock flow) play a crucial role in the explosion mechanism, and can even power observable outflows when the bounce shock fails to drive a powerful explosion \citep{Gilkis16, Quataert19, Antoni22, Antoni23}. It may be the case that these solutions are --- for some values of $n$, $\gamma$, and $\chi'_0$ --- unstable to angular perturbations as well, which would provide a physical origin for the asymmetries inferred in some CCSNe (as evidenced by, e.g., polarization in the ejecta; e.g., \citealt{Wang08, Nagao19}). Neutron star birth kicks (e.g., \citealt{Harrison75, Burrows96, Hobbs05, Scheck06, Nordhaus10, Wongwathanarat13, Janka17, Burrows24}) could also seed a relatively large $\ell = 1$ mode in the explosion, which would be amplified by the possible instability of the solutions to such asymmetric perturbations. Accounting for (small) angular deviations can be accomplished by expanding the perturbations to the fluid variables in spherical harmonics (e.g., \citealt{Ryu87, goodman90}). 

A thorough stability analysis of these solutions will be the subject of future work. 

\subsection{Implications for successful and failed explosions}
\begin{figure*}
    \includegraphics[width=0.5025\textwidth]{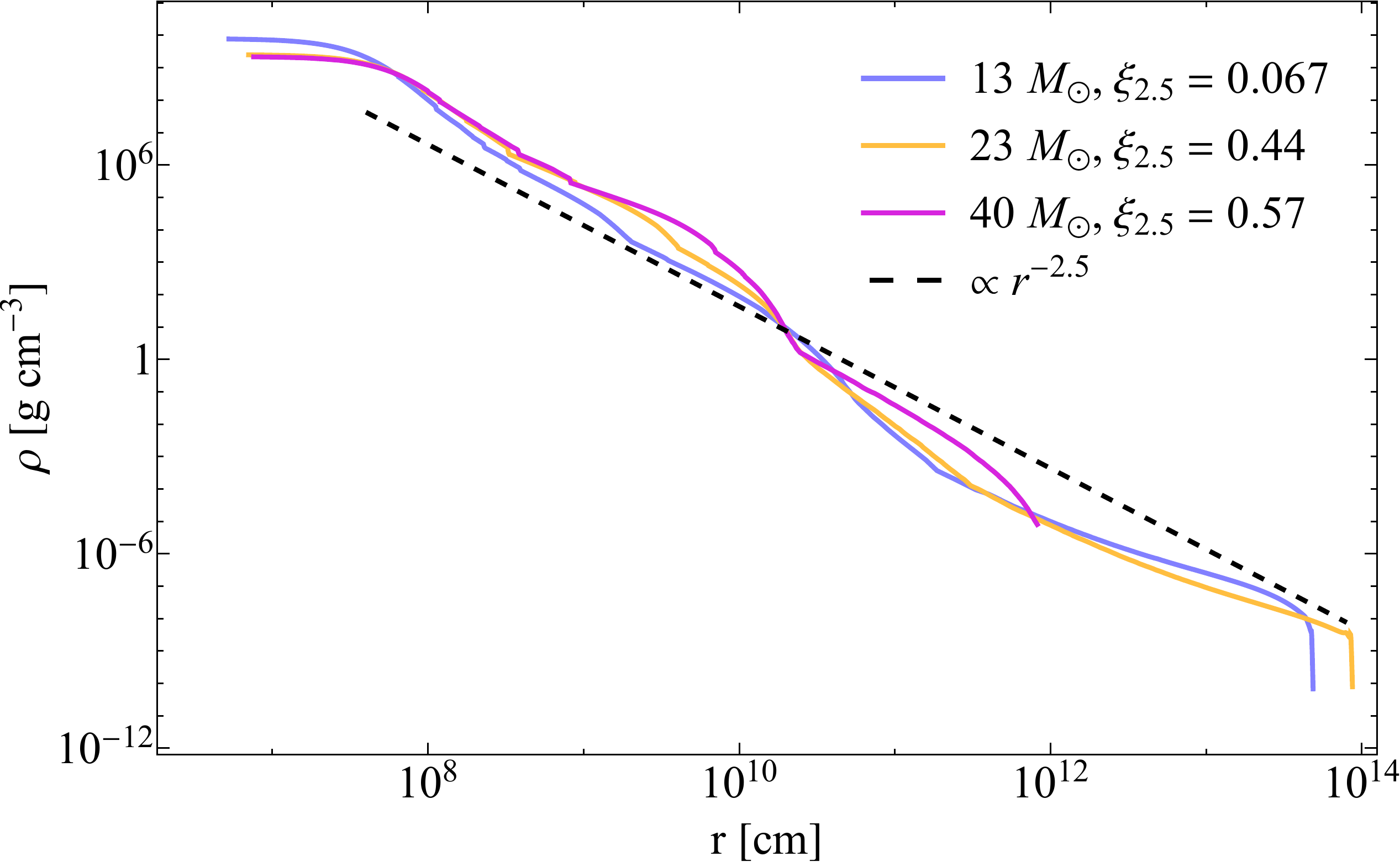}
    \includegraphics[width=0.49\textwidth]{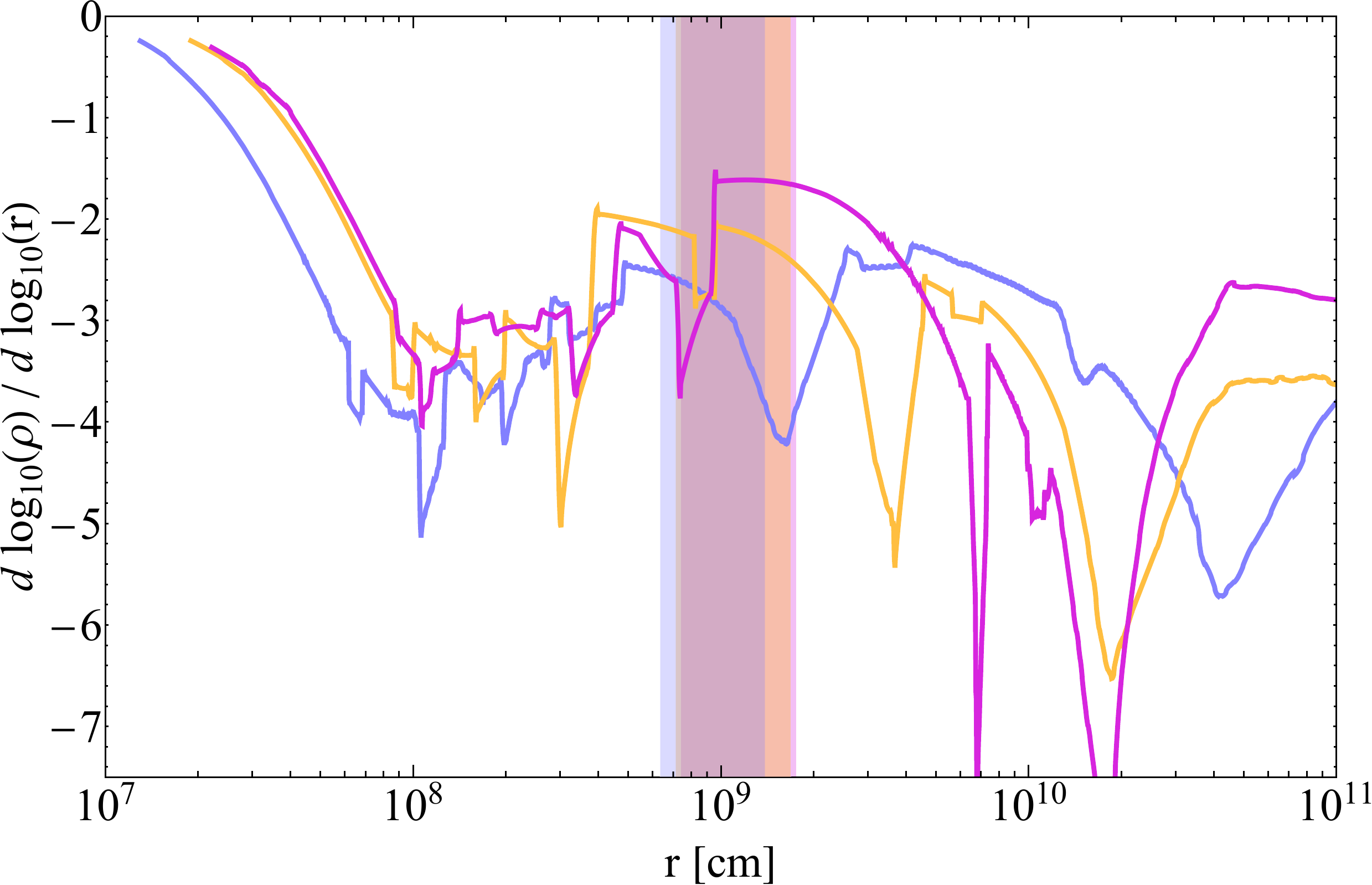}
    \caption{Left: the density as a function of radius for three different core-collapse progenitors taken from \citet{sukhbold16}, with their ZAMS masses and compactness parameters shown in the legend. The black, dashed curve indicates a $\rho \propto r^{-2.5}$ power-law. The $13~\Msun$ progenitor always produces a successful explosion (independent of explosion mechanism; see Figure 13 of \citealt{sukhbold16}), while the $23$ and $40~\Msun$ progenitors fail. Right: the instantaneous density power-law index, $d\log_{10}\rho/d\log_{10} r$, for the same stars as the left panel. The vertical shaded regions (with color corresponding to the ZAMS mass in the left plot) indicate the regions within each progenitor where the dynamical time is between $1$ and $3$ seconds, which is the neutrino diffusion time. In this region the higher-mass and higher-compactness progenitors have shallower density profiles, implying that these progenitors would require a larger velocity to produce a successful explosion.}
    \label{fig: rho_cc}
\end{figure*}
The likelihood of a massive star to produce a successful supernova explosion is correlated with its ``compactness'' parameter, which \citet{OConnor11} define as
\begin{equation}
    \xi_{2.5} = \frac{M / \Msun}{R\left(M\right) / 1000~\rm km}\bigg\lvert_{t=t_{\rm bounce}},
\end{equation}
where $M=2.5~\Msun$, and $R\left(M\right)$ is the radius that encloses $2.5~\Msun$. Stars with steep density profiles surrounding the iron core therefore have smaller compactness parameters, as they will enclose this mass at a larger radius. Alternatively, stars with shallow density profiles enclose $2.5 M_{\odot}$ at a smaller radius, leading to a higher compactness parameter, an increased accretion rate onto the newly-formed neutron star, and a higher likelihood of forming a black hole in a FSN. There are, however, drawbacks to using a single number for characterizing the structure of an evolved massive star, as the compactness parameter is degenerate with varying internal structure \citep{Ugliano12, sukhbold14, sukhbold16, sukhbold18}. Specifically, it is possible that a high central core density enclosed by a rapidly declining envelope can give similar compactness values to a lower central-density core surrounded by a region with a shallow density gradient. Nonetheless, the significance of the density profile of the region surrounding the iron core is apparent in other works as well, such as the two-parameter approach of \cite{Ertl16}, where their parameter $\mu_4$ (their Equation 3) serves as a proxy for the density gradient.

The left panel of Figure \ref{fig: rho_cc} shows the density profiles of three core-collapse progenitors taken from \citet{sukhbold16}, with their zero age main sequence (ZAMS) masses and compactness parameters, $\xi_{2.5}$, shown in the legend. Here we use the method employed by \cite{sukhbold14}, which is to calculate the compactness parameter at the onset of core-collapse instead of core bounce, which they showed has little impact on the compactness value. These three stars were chosen because \citet{sukhbold16} found that the explosion outcomes for these stars is independent of the central engine driving the explosion, with the low-mass, $13~\Msun$ progenitor always exploding and the higher mass stars always failing. The black, dashed curve shows the power-law $\rho \propto r^{-2.5}$. The right panel of Figure \ref{fig: rho_cc} shows the instantaneous power-law index of the density, $d\ln \rho/d\ln r$. 

These curves demonstrate that the $13~\Msun$ progenitor has a steeper density profile compared to the $23$ and $40~\Msun$ progenitors, which have relatively comparable power-law indices in the region immediately surrounding the core. The vertical, shaded columns --- which are colored corresponding to each progenitor (note that the $23$ and $40~\Msun$ regions are nearly the same) --- indicate the regions within each progenitor where the dynamical time is between $1-3$s which, following the arguments put forth in Section \ref{sec:ambient}, is approximately the radius that the RW reaches within the neutrino diffusion time and simultaneously when the shock is revived. While this radius is comparable for each progenitor ($\sim 10^9$cm), the respective power-law indices vary significantly, with the $13$, $23$, and $40~\Msun$ progenitors having an average instantaneous power-law index of $n\sim2.9$, $2.3$, and $2.0$, respectively, within this region. Our analysis demonstrates that shallow power-law indices require larger shock speeds for successful explosions, such that the $23$ and $40~\Msun$ progenitors would require a larger shock velocity to unbind the star, and in general should be more difficult to explode. Our results are thus generally consistent with the fact that these two progenitors are more prone to failure. 

\section{Summary and Conclusions}
\label{sec: sum}
We derived new self-similar solutions that describe the propagation of a shockwave into an infalling and time-dependent ambient medium, the density profile of which conforms to a power-law ($\rho \propto r^{-n}$ with $n > 2$) at large radii. The solutions are characterized by a sonic point in their interior, with the material approaching supersonic freefall onto a point mass at the origin, making them applicable to the early phases of CCSN when the shock is out of causal contact with the neutron star and/or after the neutron star collapses to a black hole. The shock speed is a fixed fraction of the local freefall speed, and we interpret these solutions as fluid generalizations of marginally bound orbits, such that the shock speed in an explosion must be greater than the critical shock speed derived here (and shown in Figures \ref{fig:Vcrit_n} and \ref{fig:Vcrit_n_gamma}).
 
The fact that these solutions maintain accretion onto a compact object makes them relevant to failed supernovae (FSNe), where a black hole is formed following the implosion of the neutron star. To date, there have been two detections of possible FSN candidates --- one being a red supergiant\footnote{Or possibly a yellow hypergiant \citep{Humphreys19}.} with a ZAMS mass of $\sim 25~\rm M_{\odot}$ (NGC 6946-BH1; \citealt{Adams17}) and the other a hydrogen-depleted supergiant with a ZAMS mass of $\sim 20~\rm M_{\odot}$ (M31-2014-DS1; \citealt{De24}) --- however, the interpretation of the former event is still contested \citep{Beasor24, Kochanek24}. Large sky surveys such as the Rubin Observatory Legacy Survey of Space and Time \citep{Ivezic19} will present an opportunity to increase the number of FSN candidates \citep{Byrne22}, thereby broadening our understanding of the terminal stage of massive stellar evolution. 
\newline
\newline
We thank the anonymous referee for a useful and constructive report that added to the quality and clarity of the paper. D.A.P.~thanks Ananya Bandopadhyay for useful discussions. D.A.P.~acknowledges support from the National Science Foundation through the Graduate Research Fellowship Program under grant No.\ CON05112. Any opinions, findings, and conclusions or recommendations expressed in this material are those of the authors and do not necessarily reflect the views of the National Science Foundation. E.R.C.~and D.A.P.~acknowledge support from NASA through the Astrophysics Theory Program, grant 80NSSC24K0897.

The software used in this work was developed in part by the DOE NNSA- and DOE Office of Science-
supported Flash Center for Computational Science at the University of Chicago and the University of
Rochester.
\software{FLASH \citep{fryxell00}}

\bibliographystyle{aasjournal}
\bibliography{ref}

\end{document}